\documentclass[acmsmall]{acmart}

\AtBeginDocument{%
  \providecommand\BibTeX{{%
    \normalfont B\kern-0.5em{\scshape i\kern-0.25em b}\kern-0.8em\TeX}}}

\setcopyright{acmcopyright}
\acmJournal{TWEB}
\acmYear{2021} \acmVolume{1} \acmNumber{1} \acmArticle{1} \acmMonth{1} \acmPrice{15.00}\acmDOI{10.1145/3452332}

\graphicspath{ {./images/} }
\usepackage{booktabs}
\usepackage{amsmath}
\usepackage{algorithm}
\usepackage{algorithmic}

\usepackage{amssymb}
\usepackage{multirow}
\usepackage{enumitem}

\usepackage[utf8]{inputenc}
\usepackage[english]{babel}
\usepackage{amsthm}
\theoremstyle{definition}

\usepackage{xcolor}
\usepackage{enumitem}

\usepackage{array}

\usepackage{subfig}

\begin{document}

\title{Sequential Learning-based IaaS Composition}


\author{Sajib Mistry}
\affiliation{%
  \institution{School of Elec Eng, Computing and Mathematical Sciences, Curtin University}
  \state{WA}
  \postcode{6102}
  \country{Australia}}
\email{sajib.mistry@curtin.edu.au}

\author{Sheik Mohammad Mostakim Fattah}
\affiliation{%
  \institution{School of Computer Science, University of Sydney}
  \state{NSW}
  \postcode{2006}
  \country{Australia}}
\email{sheik.fattah@sydney.edu.au}

\author{Athman Bouguettaya}
\affiliation{%
  \institution{School of Computer Science, University of Sydney}
  \state{NSW}
  \postcode{2006}
  \country{Australia}}
\email{athman.bouguettaya@sydney.edu.au}

\renewcommand\shortauthors{S. Mistry. et al}

\begin{abstract}
We propose a novel IaaS composition framework that selects an optimal set of consumer requests according to the provider's qualitative preferences on long-term service provisions. Decision variables are included in the temporal conditional preference networks (TempCP-net) to represent qualitative preferences for both short-term and long-term consumers. The global preference ranking of a set of requests is computed using a \textit{k}-d tree indexing based temporal similarity measure approach. We propose an extended three-dimensional Q-learning approach to maximize the global preference ranking. We design the on-policy based sequential selection learning approach that applies the length of request to accept or reject requests in a composition. The proposed on-policy based learning method reuses historical experiences or policies of sequential optimization using an agglomerative clustering approach. Experimental results prove the feasibility of the proposed framework. 
\end{abstract}



\maketitle

\section{Introduction}

%
%
%
%

\label{intro}

\textit{Cloud computing} is inexorably becoming the technology of choice among big and small businesses to deploy and manage their IT infrastructures and applications \cite{wang2016spatial}. Infrastructure-as-a-Service (\textit{IaaS}) is a key cloud service delivery model. Large companies such as Amazon, Google, Microsoft, and IBM provide IaaS solutions to their consumers. Examples of IaaS consumers include Software-as-a-Service (SaaS) providers and organizations such as governments, universities, and research centers \cite{fattah2020long}. The computing resources or Virtual Machine (VM) instances are the most common IaaS services \cite{hwang2015cloud}. The \textit{functional} properties of a VM instance or an IaaS service include computing resources such as CPU units, memory units, storage, and network bandwidths. Examples of \textit{non-functional} properties or Quality of Service (QoS) attributes of IaaS services are availability, price, response time, throughput, and energy efficiency \cite{fattah2020event,fattah2020signature}.

The \textit{market-driven} cloud service provisioning is a topical research area \cite{varghese2018next}. There exist several key service \textit{provisioning} models in the cloud market such as on-demand, reservation \cite{chaisiri2011optimization, zheng2016probabilistic}, and economic models \cite{sharma2012pricing,Thanakornworakij,PAL2013113}. We propose an \textit{alternative strategy to the on-demand model} which would be used in conjunction. The premise is that the on-demand or reservation model makes it difficult to accurately predict service demand, thus potentially leading to either under-provisioning or over-provisioning \cite{dustdar2011principles,jiang2011asap, islam2012empirical}. 
The alternative model focuses on the economic model-based cloud service selection and provision for long-term IaaS composition. In that regard, the economic model-based service provisioning approach is fundamentally different from the typical on-demand and reservation models. According to the \textit{on-demand}, the provider has a fixed set of VM instances associated with QoS and price \cite{Hong}. The consumer may acquire and release on-demand VM instances anytime and only pay for the usage by per hour or per second. 
The provider usually sets a discounted flat rate for the reserved instances in the \textit{reservation} model \cite{fattah2019long}. The consumers reserve the VM instance for a fixed time period and pay for it regardless of usage. 
According to \textit{economic model} based service provisioning approaches, there exists \textit{market competitions} among providers to set the price and QoS of their services \cite{PAL2013113}. The market competition forms \textit{non-cooperative games} among competitive providers and consumers \cite{Thanakornworakij}.


\textit{We focus on the economic model-based cloud service selection and provision for a long-term period}. Economic expectations are \textit{formally} expressed in terms of economic models \cite{ye2014economic,sajibtsc2015,goiri2012economic}. According to the economic model-based service selection approaches \cite{ye2014economic,yu2007efficient,kholidy2014qos}, a consumer's requests include custom VM instances, QoS parameters and the price the consumer is willing to pay for the instance (usually determined by their market research or the \textit{consumer economic models}). Similarly, the providers follow their own economic model to \textit{accept} or \textit{reject} the requests from the consumers \cite{sajibtsc2015}. 
The economic model-based service selection and provisioning approaches are applied in different cloud markets such as spot market, SLA negotiation, and auction-based reservation models \cite{zaman2013combinatorial}.

We assume that consumers create their \textit{IaaS requests} following their own economic models. An IaaS provider receives a set of IaaS requests from \textit{different} consumers. \textit{The IaaS composition from the provider's perspective is defined as the selection of an optimal set of consumer requests \cite{sajibtsc2015}}. The IaaS composition is a \textit{decision-making} problem where the provider decides which requests it should \textit{accept} or \textit{reject}. An effective IaaS composition \textit{maximizes} the provider's long-term economic expectations, such as profit, reputation, and consumer growth \cite{fattah2018cp}. The economic model is the \textit{formal tool} to select the optimal set of consumer requests to meet the provider's expectations \cite{dash2009economic}. 

Our objective is to design a \textit{qualitative economic model} for the long-term IaaS composition \cite{sajibicsoc2016}. The qualitative economic models provide an \textit{effective} way to select consumer requests where there exists \textit{uncertainties} or \textit{incomplete} information. The consumer requirements are typically uncertain and probabilistic \cite{fattah2018cp} for the long-term period. \textit{The provider's long-term economic expectations are also dynamic} \cite{mistry2018long}. The qualitative economic model specifies the provider's \textit{temporal business strategies} such as reputation building, risk management, revenue, and profit maximization \cite{sajibicsoc2016}. These business strategies determine the service provisioning \textit{preferences}. For example, the provider may observe very high demand for \textit{Network-intensive services} (i.e., VM instances designed for Network-intensive applications, e.g., C5n instance type in Amazon EC2\footnote{https://aws.amazon.com/ec2/instance-types/c5/}) in the Christmas or holiday period. The provider may prefer to provision Network-intensive services than CPU-intensive services (i.e., VM instances designed for CPU intensive applications, e.g., P3 instance in Amazon EC2\footnote{https://aws.amazon.com/ec2/instance-types/p3/}) to increase its revenue. 

We assume that an IaaS provider has its long-term qualitative economic model, i.e., the temporal service provisioning \textit{preferences} \cite{sajibicsoc2016}. The provider receives long-term IaaS requests from different consumers which are represented in \textit{time series} and associated with QoS parameters and price. The \textit{qualitative IaaS composition} is defined as the \textit{selection} or \textit{acceptance} of an optimal set of IaaS requests using the qualitative economic model of the provider. \textit{We aim to provide a comprehensive framework for long-term qualitative IaaS composition}. To the best of our knowledge, apart from our previous work \cite{sajibicsoc2016,mistry2018long}, existing research mainly focus on the \textit{quantitative IaaS composition}. The target of the quantitative composition is to \textit{maximize revenue and profit} of the provider for a \textit{short-term} period without any long-term business strategies or economic model \cite{ye2013qos,chaisiri2012optimization,zhu2010resource}. In contrast, the target of the qualitative composition is to \textit{maximize the similarity measure} between a given set of consumer requests and the provider's qualitative economic model.

We represent the provider's long-term qualitative economic model using \textit{Temporal Conditional Preference Networks} (TempCP-nets) \cite{mistry2017probabilistic}. The TempCP-net \textit{ranks} the short-term and long-term consumer requests using k-d tree according to provider's preferences \cite{sajibtsc2014}. The qualitative composition is \textit{transformed} into a \textit{combinatorial optimization problem} where the objective is to select the consumer requests that maximizes the preference rankings. We explore two composition approaches: a) \textit{global composition}, and b) \textit{local composition} \cite{alrifai2009combining, yu2008framework}. The global composition approach considers all the consumer requests within the composition interval which is computationally expensive \cite{sajibicsoc2016}. The local composition approach \textit{divides} the composition interval into several time segments and optimizes a \textit{partial set} of requests (acceptance or rejection) in each time segment. It may \textit{significantly improve} the runtime efficiency as we do not need to consider the whole set of requests during the entire composition period. However, the local composition approach is a \textit{greedy approach} of sequential optimization \cite{gnanlet2009sequential,pednault2002sequential} and may not produce the optimal result as the request selection is \textit{temporal-dependent} on the previous acceptance or rejection decisions in other time segments. For example, when we optimize the requests from left to right time segments (i.e., January, February, March), the composition result may be different than the optimization from right to left time segments (i.e., March, February, and January). A reinforcement learning based approach called 3d Q-learning \cite{mistry2018long} is proposed to \textit{find the optimal sequence} of temporal selections. The proposed 3d Q-learning based composition approach is considered \textit{off-policy} as the \textit{learning approach has no restrictions over exploration}. The proposed approach does not consider the \textit{temporal distribution} and \textit{correlations} of the \textit{historical request sets} to compose a new set of requests using \textit{policy reuse}.  


We propose a novel \textit{on-policy} based 3d Q-learning approach that effectively utilizes historical information to find the optimal selection of requests. First, the proposed learning approach reduces the run-time by removing redundant state transitions in the off-policy based 3d Q-learning approaches. Next, a novel request annotation approach based on agglomerative clustering techniques \cite{fernandez2008solving,bouguettaya2015efficient} is proposed to capture the \textit{intrinsic characteristics} of the historical requests such as the temporal distribution and the global preference ranking. A \textit{novel policy reuse approach} is proposed to compose a new set of requests that effectively utilizes previous policies which are learned from historical information. The key contributions of this work are as follows:

\begin{itemize}
    \item A comprehensive framework to compose long-term and short-term IaaS requests based on the provider's qualitative economic model.
    \item An on-policy based 3d Q-learning approach to finding the optimal request selection sequence.
    \item A novel request annotation approach to capture the intrinsic characteristics of historical request sets.
    \item A novel policy reuse approach to enable effective utilization of historical information in the proposed 3d Q-learning.  
\end{itemize}

The rest of the paper is structured as follows. In section 2, we introduce a set of terminologies and concepts that are used to formulate the qualitative composition problem. Section 3 illustrates a motivation scenario to explain the need for sequential learning in IaaS composition. Section 4 provides a general overview of the proposed qualitative IaaS composition framework. In section 5, we describe the proposed IaaS composition approach for a new set of requests. Section 6 describes the proposed long-term qualitative composition with the previous learning experience. Section 7 presents our experiments to evaluate the proposed approaches. Section 8 summarizes the related work on economic model based cloud service composition and sequential learning approaches. Finally, Section 9 concludes the paper and discusses the limitation and future work of this paper.  

\section{Preliminaries}

In this section, We introduce a set of terminologies and concepts that are used to formulate the qualitative IaaS composition problem in this paper. We will use these terminologies throughout the paper to describe the problem and proposed solution.

\begin{itemize}
    \item \textit{Providers and Services:} IaaS providers are referred as the provider. We consider the composition problem from the perspective of a single provider. IaaS services usually include a wide selection of VM instance types optimized to fit different use cases such as general purpose, compute optimized, memory optimized, and storage optimized \cite{hwang2015cloud}. The VM Instance types comprise varying combinations of CPU, memory, storage, and networking capacity and give consumers the flexibility to choose the appropriate mix of resources for their applications. We consider VM instances as services. The VM instances designed for Memory-intensive applications are termed as Memory-intensive services. For example, The R5 instance in Amazon EC2 is an example of Memory-intensive services\footnote{https://aws.amazon.com/ec2/instance-types/r5/}. Similarly, VM instances designed for CPU-intensive applications are termed as CPU-intensive services. For example, the P3 instance in Amazon EC2 is an example of Memory-intensive services\footnote{ https://aws.amazon.com/ec2/instance-types/p3/}. 
    
    \item \textit{Resources:} The resources are the capacity of the physical machines that are used to offer the VM services such as: CPU cores, memory unit, and network bandwidths \cite{hwang2015cloud}. \textit{We assume that provider has a fixed set of resources.} Here, the number of fixed set of resources refers to the maximum number of resource the provider may have at a certain point of time. The maximum number of resource however may change over period of time. In such a case, the number of fixed of resources is required to be updated according to the new maximum number of resources. The proposed approach can be considered as a proactive approach, where provider anticipates the maximum number of resources it can have.  Our aim to utilize these resources based on the provider's economic model.
    
    \item \textit{Consumers}: The targeted consumers are mainly the medium to large business organizations such as SaaS providers, governments, universities, and research institutes. These organizations may require services for a long-term period (e.g., 1 to 3 years).  
    
    \item \textit{IaaS Requests}: IaaS requests refer to the configuration of \textit{functional} and \textit{non-functional} requirements of the VM over a period of time. A consumer may need a VM with 2 vCPU, 2gb memory, and 99\% availability in first six months of a year. The IaaS requests for that period is represented as (2 vCPU, 2 gb memory, 99\% availability). \textit{We assume the deterministic IaaS requests, i.e., the provider has knowledge about the long-term requests prior to the composition}. We compose these requests based on the provider's economic models. The provider defines the requests as either \textit{short-term} or \textit{long-term}. A request is considered short-term if only one business strategy is applicable for the life time of the requests. A request is considered long-term if more than one business strategies are applicable to the request. For instance, if the provider changes its business strategies or economic models quarterly, a request that needs reservation for a VM for 1 year is considered as a long-term request. In this circumstances, a request that reserves the VM for a one month is considered as short-term requests. Note that, we focus on the economic model-based service provisioning where VMs are reserved for a certain time period. \textit{The burstable on-demand resources are outside the focus of this paper}. 
    
    \item \textit{Conditional Preference Networks (CP-nets)}: The CP-net is a widely used tool that captures a user's conditional preferences qualitatively \cite{cp1}. CP-Nets \cite{boutilier2004cp} is a compact and intuitive formalism for representing and reasoning with conditional preferences under the ceteris paribus (``all else being equal") semantics. The dynamic semantics of the preferences are indicated using a Conditional Preference Table (CPT) \cite{boutilier2004cp}. One CP-net can only represent one business strategy at a time \cite{sajibicsoc2016}. For example, if the business strategy is to build reputation for the first three months, a CP-net could be constructed that can graphically express the preference on higher QoS in a service than higher prices. 
    \item \textit{Temporal Conditional Preference Networks (TempCP-nets)}: The TempCP-net is a set of CP-nets that represents a provider's economic expectations over the long-term period \cite{sajibicsoc2016}. If there are three business strategies for a year, the TempCP-net could be constructed with a set of three CP-nets.
    
    \item \textit{k-d Tree}: The induced graph in a CP-net may contain nodes with multi-dimensional tuples and the annotated ranking of preferences \cite{boutilier2004cp}. The \textit{k-d tree} is a graph indexing technique in which every node is a \textit{k}-dimensional point \cite{andoni2006near}. Every non-leaf node in a k-d tree can be thought of as implicitly generating a splitting hyperplane that divides the space into two parts, known as half-spaces. Points on the left and right sides of this hyperplane are represented by the left and right subtree of that node respectively. We apply the k-d tree to index the service preference rankings based on the provider's TempCP-net.
    
    \item \textit{Local IaaS Ranking}: The rank of an IaaS request is determined by its k-d tree index. The ranking of a request or a set of request considering only one period is called local IaaS ranking. For example, if an IaaS request expands from January to December, then its local IaaS ranking of January is computed considering its preference rankings on January.
    
    \item \textit{Global IaaS Ranking}: The ranking of a long-term request or set of requests considering each period is called its global IaaS ranking. For example, if an IaaS request expands from January to December, then its global ranking considers is the aggregated local ranking of each month from January to December. 

    \item \textit{Sequential Optimization}: The sequential optimization approach is a series of local optimization where the initial problem is divided into sub-problems \cite{gnanlet2009sequential}. The local optimizations have cascading effect as the decisions in each local optimization affects the sequential decision making in successive local optimizations. The key benefit of the sequential optimization is that it reduces the search space of the global optimization significantly \cite{pednault2002sequential}.    
    
    \item \textit{Q-learning}: Q-learning \cite{watkins1992q} is a model-free reinforcement learning approach. The goal of Q-learning is to learn a policy, which tells an agent what action to take under what circumstances. It does not require a model of the environment, and it can handle problems with stochastic transitions and rewards, without requiring adaptations \cite{watkins1992q}. In the context of IaaS compositions, the Q-learning is used to learn the optimal sequences of request selection. The two-dimensional Q-learning has no start and terminal states as it accepts only model-free state transitions \cite{shani2005mdp}. A Q-learning process is termed as \textit{off-policy} if the learning approach has no restrictions over exploration \cite{munos2016safe}. In contrast, the \textit{on-policy} based Q-learning process is \textit{smart} and removes redundant state transitions considering historical information \cite{van2016deep}.
\end{itemize}

\section{Motivation Scenario}

\begin{figure}
    \centering
      \includegraphics[width = .8\textwidth]{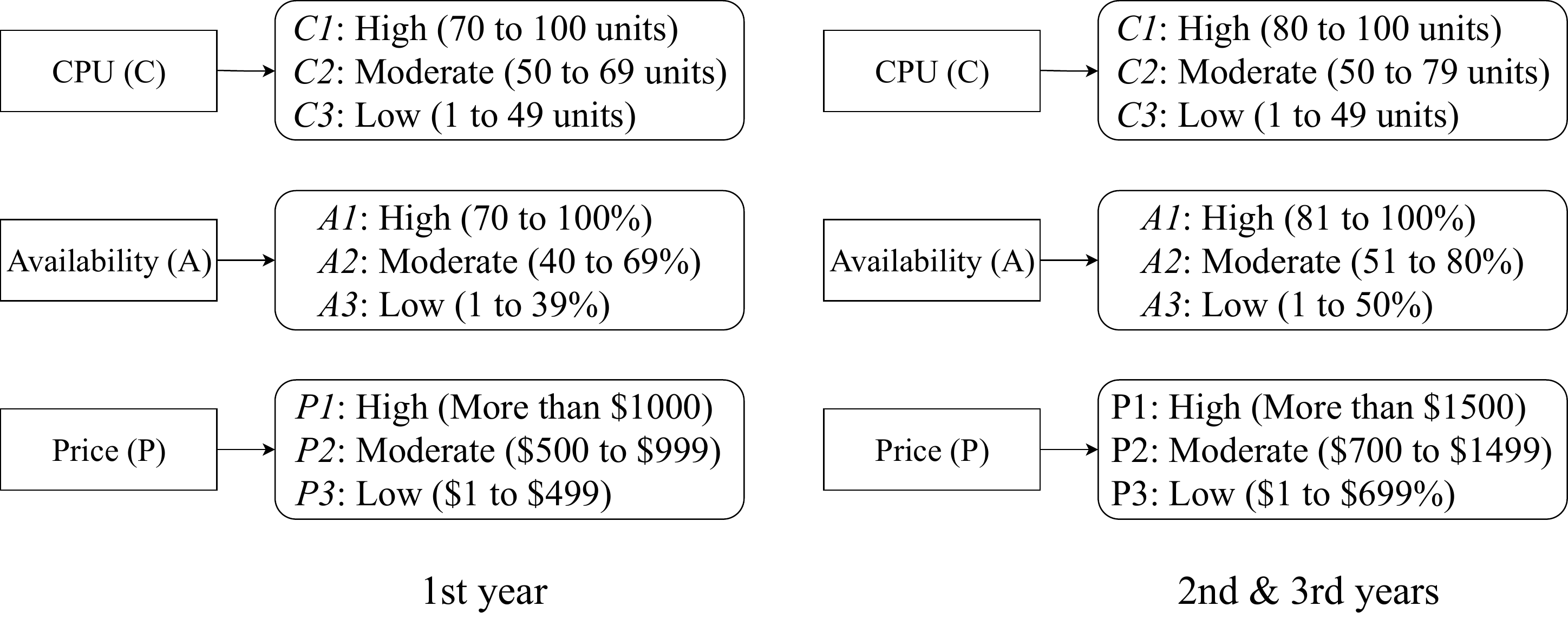}
    \caption{Semantic preference table}
       \label{fig:semanticTable}
       \vspace{-4mm}
\end{figure}

In this section, We illustrate an example scenario to describe the need of sequential learning in IaaS composition.  Let us consider an IaaS provider offers virtual CPU services. The provider offers 100 CPU units with 100\% availability. We are only considering availability as a QoS parameter for simplicity. We defined three semantic levels - high, moderate, and low to express qualitative preferences of the provider for services attributes  as shown in Figure \ref{fig:semanticTable}. We assume the provider may change the interpretation of the semantic levels based on the cloud market condition. The provider considers more than \$1000 as high price according to Figure \ref{fig:semanticTable} in the first year. The provider considers more than \$1300 as high price due to predicted inflation in the second and third years. Three different preference rankings are set based on the provider's annual goal in three years.

\begin{figure}
    \centering
      \includegraphics[width = 0.9\textwidth]{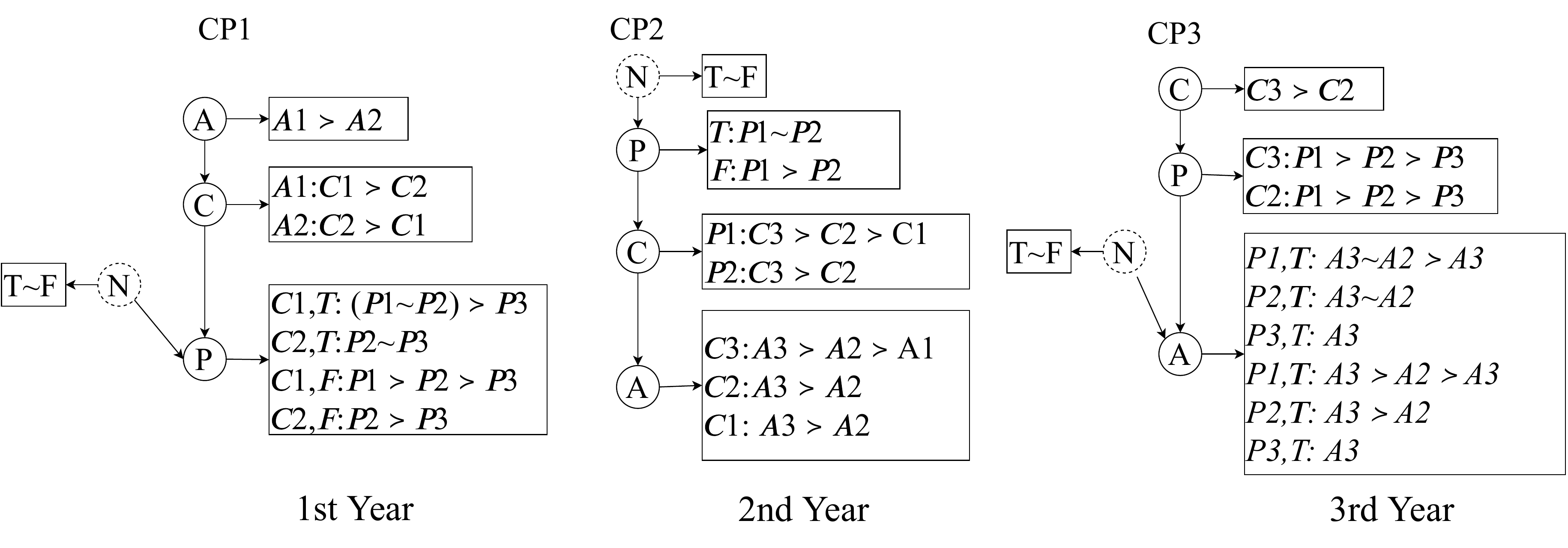} 
    \caption{CP-nets of a provider}
    \label{fig:cp}
    \vspace{-4mm}
\end{figure}

We adopt the economic models of the provider as described in \cite{mistry2018economic} to continue in this example. Figure \ref{fig:cp} shows the provider's three economic models for three different years. In the first year, the provider wants to offer high-quality service in with a lower price to create its reputation. The most important attribute in the first year is the ``availability'' of a service followed by the ``CPU'' and the ``price''. The provider decides to maximize its profit by offering services at a higher price for lower resources and QoS in the second year. The ``price'' therefore sets the ``CPU'' and the ``availability'' in the second year. The provider's preference for the third year is to provision lower CPU-intensive services. Let us assume the provider wants to receive requests that are long-term. Therefore, the provider offers discounts on long-term service requests. A decision variable labeled $N$ is used to distinguish the type of requests. The value of $N$ is set to true ($T$) when a request is long-term. A request is considered long-term if it spans over the next period. Otherwise, the value of $N$ is set to false ($F$) to indicate the request as a short-term request. In figure \ref{fig:cp}, $N$ is associated with ``price'' ($P$) for the first two years. In these periods ($CP1$ and $CP2$), the provider considers the high and moderate ``price'' level indifferently for long-term requests. $N$ is associated with ``availability'' ($A$) in the third year. According to $CP3$, short-term requests are provided with relatively lower ``availability" at the same moderate price. More details about how to represent these economic models using CP-nets can be found in \cite{sajibicsoc2016}.


\begin{figure}
    \centering
      \includegraphics[width= 0.8\textwidth]{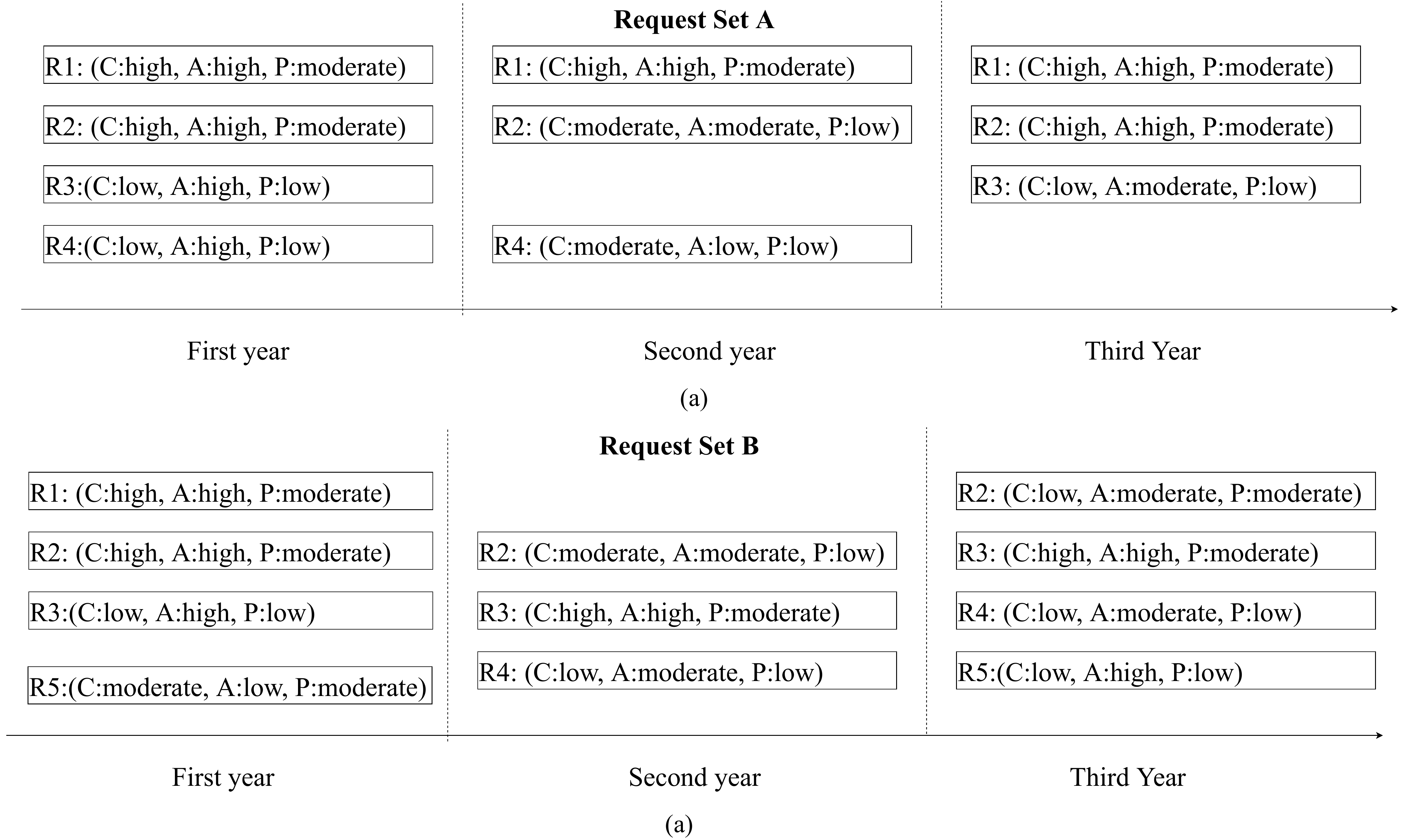}  
    \caption{ Sets of requests (a) request set $A$  (c) request set $B$}
    \label{fig:req}
    \vspace{-5mm}
\end{figure}

Let us assume a set of requests is represented by $A$ in Figure \ref{fig:req}(a). $A$ has four requests i.e.,  $\{R1\}$,$\{R2\}$,$\{R3\}$, and  $\{R4\}$ as shown in Figure \ref{fig:req}(a). Each of these requests arrives at the beginning of the composition. A request is represented in annual segments for simplicity. For instance, $(C: High, A: low, P: moderate)$ represents a request segment of $\{R1\}$ in the first year. Similarly, Figure \ref{fig:req}(a) shows the annual requirements of other consumer requests $\{R2\}, \{R3\} $ and $ \{R4\}$(a) for three years. The provider can select these four requests from $2^{4} = 16$ possible combinations to find the optimal composition in a brute force manner.

The number of possible ways to select the requests grows exponentially with the number of requests. A sequential optimization process may be applied to reduce the total number of comparisons to find the global optimal composition. Let us consider a sequential optimization approach for the request set $A$ where requests are selected from the right to left year i.e., $3^{\text{rd}}$,$2^{\text{nd}}$ and $1^{\text{st}}$. There are $2^3 = 8$ comparisons are required in the third year to select the highest ranked $R3$ according to the $CP3$. Note that, in $CP3$ the highest preference order is low CPU, high price, and low availability. In $R3$, the consumer's preference order is low CPU, moderate availability, and low price. Therefore, $R3$ is the closest match request according to $CP3$. Details of the ranking technique using CP-nets can be found in \cite{sajibicsoc2016}. Once we only accept  $R3$ in the third year, $R1$ and $R2$ are rejected in the subsequent years. In following years, the local optimization accepts $R4$ and update the solution. The optimal solution $\{R3, R4\}$ is calculated in ten comparisons. If we change the sequence of optimization process e.g., left to right i.e., $1^{\text{st}}$, $2^{\text{nd}}$ and $3^{\text{rd}}$, the total number of comparison in the first year becomes $2^4 = 16$ to select the highest ranked $R1$. The request $R1$ has ``N/A" ranking in the following years. As a result, the left to right sequence produces an unacceptable solution. The right to left sequence generates optimal result when sequential optimization is applied on $A$. The same sequence may not work or give a good solution for a different set of requests. Let us consider the request set $B$ in Figure \ref{fig:req}(c). The distribution of requests in $B$ is different from $A$. The number of comparisons becomes $2^5 = 32$  to select the highest ranked $R3$ in the third year (Figure \ref{fig:req}(d)). As $R3$ has ``N/A" ranking in the second year, it can not be selected. As a result, the right to left sequence does not work for $B$. We propose a model-free learning approach to generate the optimal sequence of local optimizations.

\section{A Qualitative IaaS Composition Framework}

In this section, we provide a general overview of the proposed qualitative IaaS composition framework. We also describe some common concepts in qualitative IaaS composition such as long-term IaaS request representation, long-term qualitative preferences representation, combinatorial optimization in qualitative IaaS composition.

\begin{figure}
	\centering
  	\includegraphics[width= .6\textwidth]{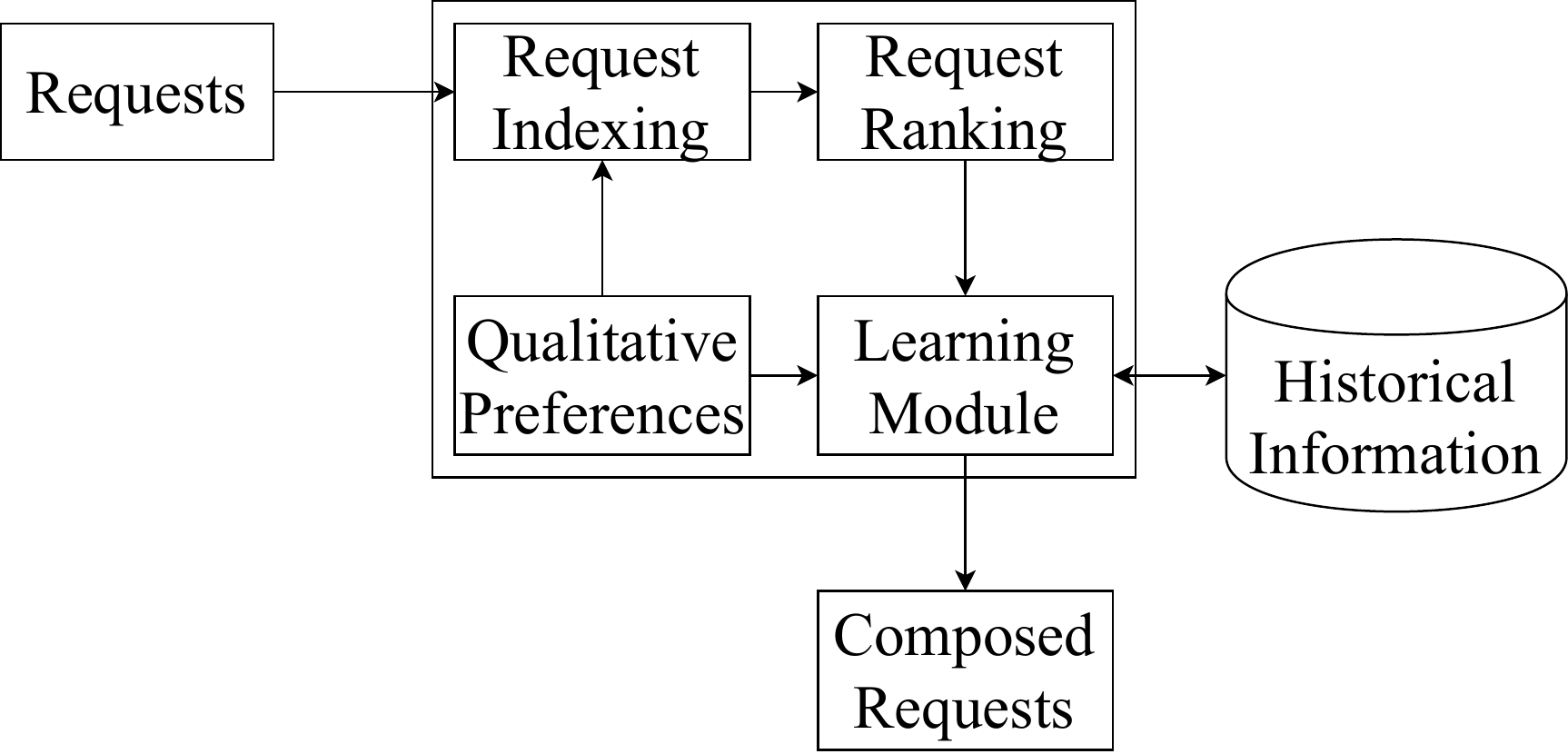}   
    \caption{Long-term qualitative composition framework}
  	\label{fig:propframe}
\end{figure}

A qualitative composition framework is proposed that learns from historical information of the past request sets as shown in Figure \ref{fig:propframe}. We assume that a set of long-term requests of consumers and the provider's qualitative preferences are available at the beginning of the composition. Our target is to find the optimal composition using a learning based approach that utilizes information of the past consumer requests. For each set of new requests, we learn the sequence of the optimal composition and save it for future request sets.

We use TempCP-nets to represent the qualitative preferences which enable qualitative composition of the requests. The proposed framework performs indexing and ranking of the requests configuration of the TempCP-net. The indexing of the TempCP-net is built using \textit{k}-d tree indexing that enables efficient searching of the rank of the consumer requests. The ranking of the requests configurations is computed using the provider's TempCP-nets. The ranked requests are taken as an input of a learning module. The learning module applies a reinforcement learning based method that utilizes the historical information of the past requests to find out optimal composition efficiently.


\subsection{\textbf{Long-term IaaS Request Representation}}

The long-term requests of the consumers are represented as time series group (TSG) of the service attributes. We denote \(T\) as the total service usage time. The TSG of the consumer requests is defined as \(R_c = \{s_{c1},s_{c2},...,s_{cn}\} \), where $s_{cn}$ represents a service attribute and \(cn\) is the number of service attributes in \(R_c\). We represent the service attribute time series as \( s_{cn} = \{(x_n,t_n)|n=1,2,3,...., T\} \), where \(x_n\) is the value of \(s_{cn}\) at the time interval \(t_n\). Figure \ref{fig:req} shows two sets of requests where each request in a set has 3 service attributes ($cn=3$) i.e., CPU, availability, and price. Each requests has three year intervals ($T =3$). Each service attribute may have different types of value during these intervals.


\subsection{\textbf{Long-term Qualitative Preferences based on TempCP-nets}}
\label{sec:cp}

We need an efficient tool to represent the provider's qualitative preferences. We define a set of attributes $V = \{X_{1},..., X_{n}\}$ which is defined over the finite, discrete domain $D(X_n)$ and semantic domains $S(X_n)$. The attributes are either functional or non-functional. Examples of functional attributes are CPU ($C$), Memory ($M$), and so on. Availability ($A$), Price ($P$), Latency ($LT$) are examples of QoS attributes. A mapping table $Sem\_Table(X_{n}, x_{n})$ is used to map $x_{n}$ in $D(X_{n})$ into $s_{n}$ in  $S(X_{n})$ where $s_{n} = Sem\_Table(X_{n}, x_{n})$. An example of a semantic table is shown in Figure \ref{fig:semanticTable}. We assume the preferences order and semantics of $V$ is static within an interval in a long-term composition period. However, they may vary within different intervals. We consider a set of decision variables $DN = \{N_{1}, N_{2},....,N_{d}\}$. A decision variable may represent request type, requests duration, and etc. We assume that the decision variable is a binary variable. Therefore, it takes true of false $\{T, F\}$ values. For instance, the decision variable is set true for a request if it spans to the next interval. 

\begin{figure}
	\centering
  	\includegraphics[width= .8\textwidth]{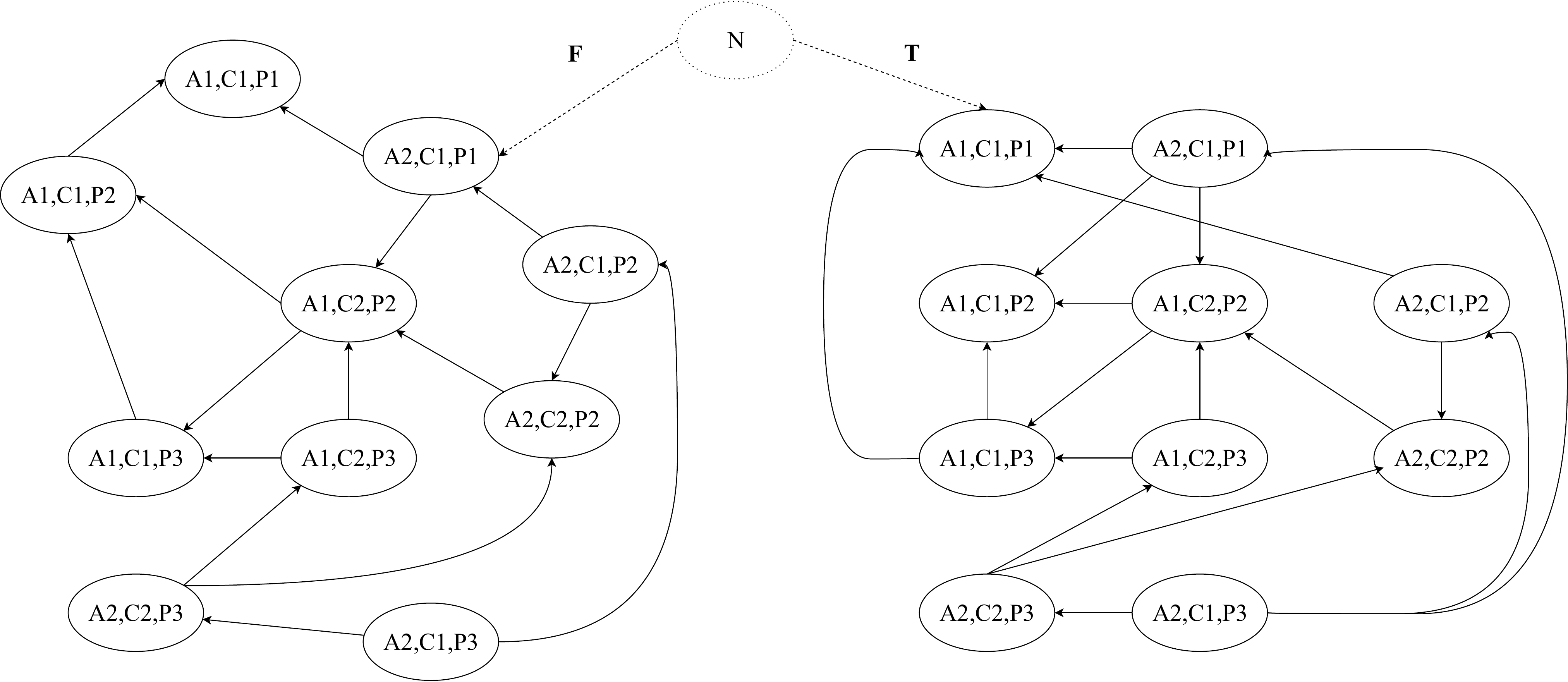}   
    \caption{Induced preference graph with decision variable for CP1}
  	
	\label{fig:f3}
	\vspace{-3mm}
\end{figure}

The service provisioning time $T$ is considered into $m$ intervals and represented as  $T=\sum_{k=1}^{m}I_{k}$, where $I_k$ is an interval in $T$. We assume that the provider sets a preference ranking of each service configuration at each interval $I_k$. The preference rankings of service configurations are expressed over the complete assignments on $V$ and $DN$ with the semantic domain $Sem\_D^{I_{k}}(V)$. $O^{I_{k}}$ is denoted as the set of service configurations for an interval $I_{k}$. A total order $(\succeq)$ of the service configuration set represents a preference ranking for an interval. For example, $o_{1} \succeq o_{2}$ denotes that a service configuration $o_{1}$ is equally or more preferred over $o_{2}$. The preference relation $o_{1} \succ o_{2}$ denotes that the service configuration $o_{1}$ is preferred over $o_{2}$. If the preferences are indifferent or non-comparable, we denote the relation using $o_{1} \sim o_{2}$. $T \sim F$ means that the provider does not care about the true and false values of a decision variable. 

The size of the service configuration set $O^{I_{k}}$ grows exponentially with the number of intervals. Therefore, direct assignment of all possible preferences over the long-term period is always not feasible. We represent the provider's long-term preferences on service configurations using a TempCP-net. A TempCP-net is represented as set of CP-nets with semantic preference tables for each interval of the composition period. We denote a TempCP-net as $\text{TempCP-Net} = \{(CP^{I_{k}}, Sem\_Table^{I_{k}}, I_{k})\;|\;\forall k \in [1,m]\}$. A CP-net can be considered as a graphical model that formally represents qualitative preferences and reasons about them. A CP-net $CP^{I_{k}}$ in the interval $I_{k}$ consists a direct graph $G$ which is defined using $V$ and $DN$. Each node in $G$ represents an attribute $X_i \in V$. The nodes of $DN$ are represented by a dashed circle. A node of $V$ is represented by a solid circle. In this work, we only consider acyclic CP-nets to represent the provider's qualitative preferences. The CPT of each node is denoted by  $CPT(X_{i})$ which contains a total order $\succ^{i}_{u}$ with each instantiation $u$ of $X_{i}$'s parents $Pa(X_{i}) = U$ \cite{cp1}. For example, $Pa(P)=C$ and $CPT(C)$ contains $\{C1, C2\}$ in $CP3$ while preferences are made over $\{P1, P2, P3\}$ (Figure \ref{fig:cp}). A preference outcome $o$ of a CP-net is obtained by sweeping through the CP-net from top to bottom setting each variable to its preferred value given the instantiation of its parents \cite{wang2012wcp}. A preference order $o \succ \acute{o}$ is called a consequence of a CP-net, if $o \succ \acute{o}$ can be obtained directly from one of the CPTs in the CP-net. For example, the fact that $(A2, C2, P2)$ is preferred to $(A2, C1, p2)$ is a direct consequences of the semantics of $CPT(C)$ in $CP1$ for the long-term requests (Figure \ref{fig:cp}). The set of consequences $o\succ \acute{o}$ creates a partial order over all possible service configurations of an acyclic CP-net.

Figure \ref{fig:f3} shows the induced preference graph generated by $CP1$ which is a directed acyclic graph (DAG). There are two induced graph generated based on the value of the decision variable $N$ (i.e., true and false). The true value of $N$ represents the induced preference graph for long-term requests. The false value of $N$ represents short-term requests. $(A1, C1, P1)$ is considered as the most preferred request for the short-term requests. There is no outgoing edge from $(A1, C1, P1)$. Similarly, $(A2, C1, P3)$ has no incoming edge because it is least preferred request configuration. There is an edge between $(A2, C1, P1)$ and $(A1, C1, P1)$. According the preference statement $A1 \succ A2$ in the $CPT(CP1)$, $(A1, C1, P1) \succ (A2, C1, P1)$. $(A1, C1, P1)$ and $(A1, C1,\\ P2)$ does not have any outgoing edge because they are the highest preferred configuration (Figure \ref{fig:f3}). The induce preference graph of a CP-net is constructed by pairwise comparison of all service configurations. The complexity for ordering queries for a TempCP-net in an interval is $O(ndq^2)$ where $n$ and $d$ is the number of attributes and decision variables respectively and the number of output configurations is $q$.

\subsection{\textbf{Combinatorial Optimization in Qualitative IaaS Composition}}

\begin{figure}[t!]
	\centering
  	\includegraphics[width= 0.8\textwidth]{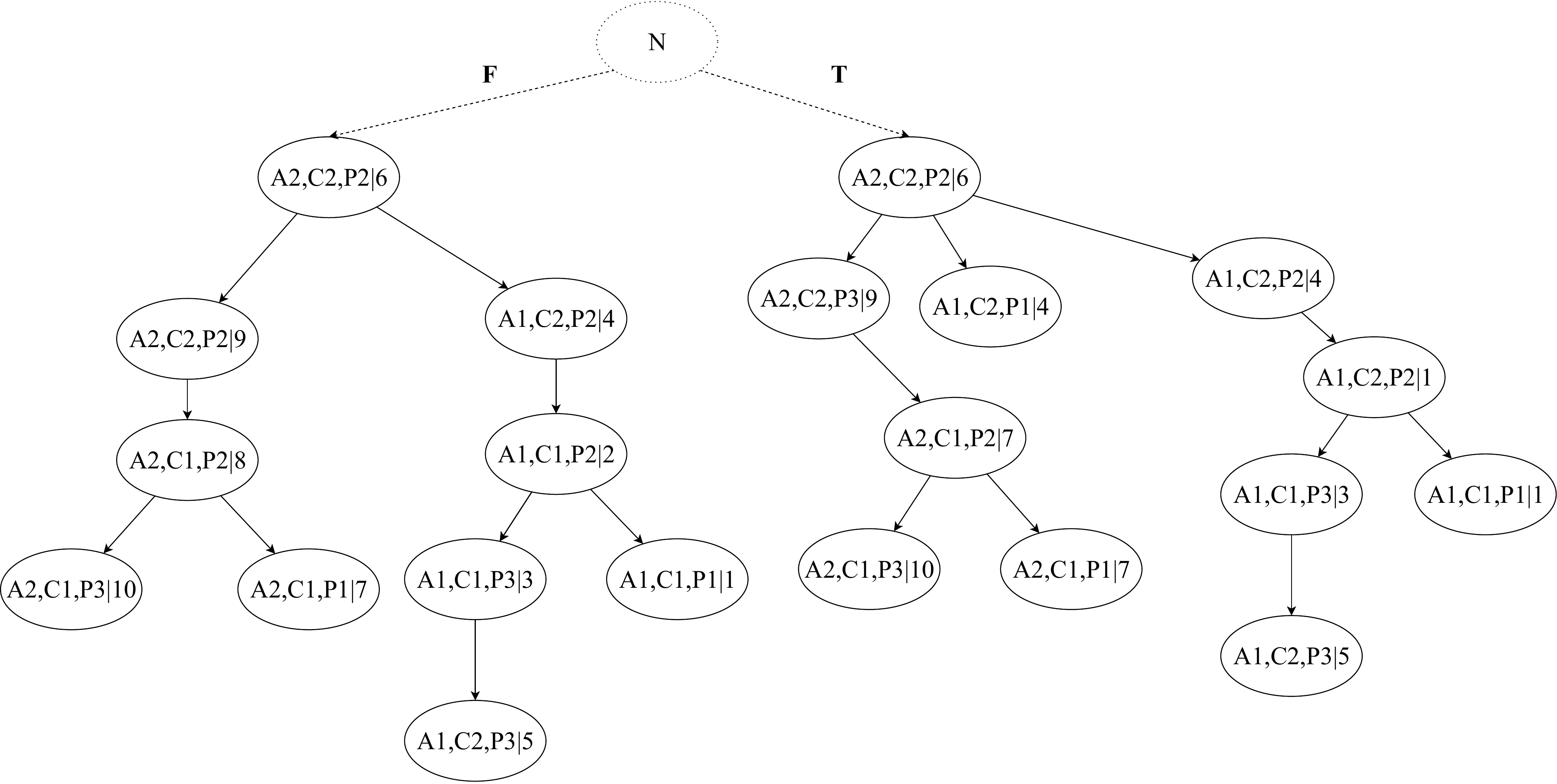}   
    \caption{\textit{k}-d tree indexing of the induced preference graphs}
    \vspace{-3mm}
	\label{fig:f4}
\end{figure}

Given a provider's TempCP-net and a set of long-term requests $R$, the IaaS composition is defined as the selection of an optimal set $\bar{r} \subseteq \bar{R}$ that produces the best similarity measure with the TempCP-Net. We consider qualitative preference rankings as the foundation of the similarity measure. First, we index preference rankings from TempCP-Net. We then perform a similarity search on the indexed TempCP-Net which is denoted as $Pref(\text{TempCP-Net}, \bar{r})$. Hence, the objective of the IaaS composition is to minimize the ranking output $Pref(\text{TempCP-Net}, \bar{r})$. 

\subsubsection{\textbf{Indexing Preference ranks}} 
\label{sec:kd}

The preference rank of a request configuration is denoted as $Sem\_Req$  $=(s_{1}, ...,s_{n}) \;|\; \text{where } s_{i} \in S(X_{i}),  \text{and }X_{i} \in V$. It is found by a pre-order traversal of the induced graph. The time complexity of searching the preference rank over the induced graph is $O(n)$. A request configuration $(s_{1}, ...,s_{n})$ is considered as a multidimensional vector. We use a \textit{k}-d tree \cite{jia2010optimizing} to improve the searching process. There exist different multi-dimensional indexing structures such as B-tree, B+-tree, k-d Trees, Point Quadtrees, R, R*, R+ Trees \cite{sellis1997multidimensional,robinson1981kdb}. The k-d tree is a \textit{congruent} choice in IaaS composition as the IaaS preference ranking requires the multi-dimensional value queries \cite{sajibicsoc2016,nam2004comparative}. \textit{Note that, finding the optimal multi-dimensional indexing structures for IaaS composition is outside the focus of this paper}.

The \textit{k}-d tree is a binary tree that is used for indexing some nodes in a space with k dimensions. We represent each service configuration $o$ (i.e., a node in the induced graph) as a k-dimensional point in the \textit{k}-d tree. Each node in each level splits its all children along a specific dimension into two subspace, known as half-spaces. Each subspace is represented by either a left or a right sub-tree of that node. A canonical method is used to build the \textit{k}-d tree \cite{jia2010optimizing}. The construction algorithm follows a cycle during the selection of splitting planes. For example, at the root, all children are split based on ``availability" plane in Figure \ref{fig:f4}. The children of the root split their children along ``CPU" plane. The grandchildren of the root have ``price-aligned" planes. Finally, the great-grandchildren have again planes aligned with availability. 

Let us assume there are $n$ points in an induced preference graph. We place the median point found in one dimension at the root of the \textit{k}-d tree. Every other point smaller and larger than the root in the same dimension are placed into right and left sub-tree respectively. This process creates a balanced \textit{k}-d tree where the runtime is $O(n \;log(n))$ \cite{jia2010optimizing}. We annotate each node of the \textit{k}-d tree with its respective preference ranking obtained from the induced graph. For instance, the root node $(A2,C2,P2)$ in the \textit{k}-d tree in Figure \ref{fig:f4} has the preference ranking 6 that is obtained from its induced graph. We construct the \textit{k}-d tree indexing for each value of the decision variable $N$. For example, two different \textit{k}-d tree indexing are shown in Figure \ref{fig:f4} to represent short-term and long-term service configurations. The service configurations with indifferent preference have the same ranking value. For example, the provider's preferences on $(A2,C1,P2)$ and $(A1,C2, P1)$ are indifferent for long-term requests. Both service configurations are annotated with preference ranking 4 in Figure \ref{fig:f4}.

\subsubsection{\textbf{Local and Global Preference Ranking}}

A request may not be inclusive, i.e., exactly fit within an interval of TemCP-net. It may overlap two or more intervals of TempCP-nets. An overlapping request $R$ (interval $[T_{0}, T_{m}]$) is divided into smaller inclusive segments where each segment fits within an interval of the TempCP-net. The attributes that have \textit{temporal semantics} require such segmentation. For instance, ``Price" is considered as an attribute with \textit{temporal semantics} in a consumer request. If a request requires 20 units of CPU for 12 months with total \$120, a monthly segmentation interprets the provisioning of 20 CPU units for \$10. Let us consider an attribute $X$ in $R$ that has \textit{temporal semantics}. If the segmentation is applied in $[T_{j}, T_{k}]$, the new segments are calculated using the following equation according to \cite{sajibicsoc2016}: 
\vspace{-3mm}

\begin{equation}
\label{eqn:rank}
x_{i}^{[T_{j},T_{k}]} =  x_{i}^{[T_{0},T_{m}]} \times \frac{|T_{k}-T_{j}|}{|T_{m}-T_{0}|}
\end{equation} 

 The requests are ready to be composed after the temporal segmentation. We define a set of $N$ requests as $\bar{R} = \sum_{i=1}^{N} R_i$. We use the following composition rules to combine the requests in a set $\bar{R}$ \cite{sajibtsc2015}:

\begin{itemize}
    \item The rule of summation: $\bar{x_{i}} = \sum_{i=1}^{N} x_{i}$, where $X_{i} \in \{C, M, NB, RT, P\}$.
    \item The rule of maximization: $\bar{y_{i}} = max(y_{i}), \forall \; i \in [1,N]$, where $Y_{i} \in \{A, TP\}$.
\end{itemize}

The preference ranking function of a set of requests is $Pref(\text{TempCP-Net}, \bar{R}): V \rightarrow [1,n]$, which outputs the order of $\bar{R}$ according to the preferences (\textit{k}-d tree of the TempCP-net). $\bar{R}$ is transformed  $\acute{\bar{R}} = \{(s_{i}, I_{j})\;|\;s_{i} \in S(X_{i}),  X_{i} \in V, \text{and } T = \sum_{j=1}^m I_{j}\}$ based on $Sem\_Table$ of the TempCP-net. First, we define the local similarity measure, i.e., preference rankings for a time segment and then we define the global objective function for the entire composition period.

\begin{itemize}
    \item \textit{Local preference ranking}: Let us consider $M^{i}(s_{i})$ in the interval $i$ is the function that outputs the preference ranking by temporal matching of  $\acute{\bar{R}}$ segments with the \textit{k}-d tree. The temporal matching process or the searching algorithm starts from the root node and traverses the tree recursively. The search algorithm returns the preference ranking of a node if that matches with a request configuration. For instance, the algorithm returns ranking 10  by performing 10 comparisons for the short-term request $(A2, C1, P3)$ in Figure \ref{fig:f4}. If a query search point is not found in the \textit{k}-d tree, it is discarded in the composition. The complexity of performing a query in a \textit{k}-d tree is on average $O(log(n))$ for each service.
    
    \item \textit{Global Preference Ranking}:  As the long-term requests are divided into local segments, we aggregate the local preference rankings to generate the global preference ranking as follows:
    \begin{equation}
    \label{eq:ranking}
        Pref(\text{TempCP-Net}, \bar{R}) = \sum_{i=1}^{m} M^{i}(s_{i})
    \end{equation}
\end{itemize}

\section{IaaS Composition for a New Set of Requests}

In this section, we illustrate the proposed IaaS composition approach for a new set of requests, i.e., IaaS composition without any prior knowledge of incoming requests. We introduce a sequential IaaS composition approach that leverages reinforcement learning techniques to compose incoming requests.

\begin{figure}[t!]
	\centering
  	\includegraphics[width=.5\textwidth]{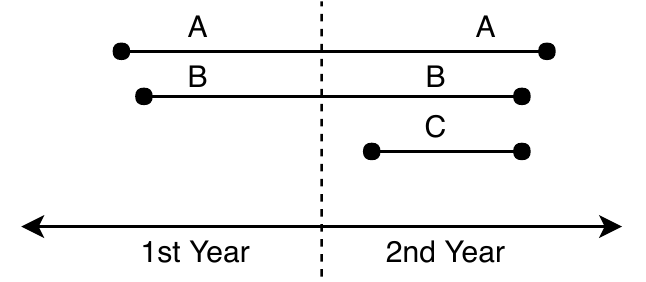}  
    \caption{Overlapping requests in different temporal segments}
    \vspace{-4mm}
	\label{fig:newExample1}
\end{figure}

We assume that initially the IaaS provider does not store a history of incoming requests. Each set of incoming requests are considered as a new set of requests and the composition is performed from scratch for the new set of requests. We identify three approaches to compose a new set of requests:
\begin{itemize}
    \item \textit{Brute-force approach}: This approach generates all the combinations of requests over the total composition period. The preference ranking of each combination is computed using the global preference ranking Equation \ref{eq:ranking} and pairwise compared. The combination of requests that generates the minimum global ranking is returned as the optimal composition. If the number of requests is $N$, the time complexity of this approach is exponential ($2^N$).
    \item \textit{Global optimization approach}: The target of the global optimization is to improve the runtime efficiency from the brute-force approach. We apply Dynamic Programming (DP) \cite{sajibicsoc2016} to reduce the re-computation of similarity measure of the same combinations of requests. The DP is designed to compute the similarity measure of a large combination of request sets by breaking it into smaller combinations (overlapping subproblem) structure. The results of the subproblems are stored in a temporary array which enables avoiding repeated computation. We denote $\bar{R}(N)$ as a set of $N$ requests and $i \in [1,N]$ as the $i_{th}$ request. $\tau(\bar{R}(N), k)$ denotes the subset of requests of size $k$ which generates the maximum preference rankings among all requests of size $k$. We start with base case $k=1$, i.e., a set consists of only one requests. The highest ranked request $i$ is computed by pairwise comparison of preference rankings:
    \vspace{-5mm}
    
    \begin{align*}
    &\text{Base case, } \tau(\bar{R}(N), k=1) = R_{i}\\ \notag &\text{where } Pref(\text{TempCP-Net}, R_{i}) \text{ is minimum.} \notag
    \end{align*}

     For $k>1$, it either accepts the $N_{th}$ request (the $k_{th}$ place is already filled) or rejects it (reduces $\bar{R}(N)$ to $\bar{R}(N-1)$). We have two optimal substructures:
     
     \vspace{-6mm}
    
    \begin{align*}
        \bar{R_{i}} &= \{N \cup \tau(\bar{R}(N-1), k-1)\} \\
        \bar{R_{j}} &= \tau(\bar{R}(N-1), k) \notag
    \end{align*}

     The $\bar{R_{i}}$ and $\bar{R_{j}}$ are computed separately. The re-computation of overlapping substructures is avoided by building a temporary array in a bottom-up manner \cite{kimes2004restaurant}. if $\bar{R_{i}}$ returns the minimum preference ranking, it should be returned as the optimal composition,  $\tau(\bar{R}(N), k) =\bar{R_{i}}$ if $Pref(\text{TempCP-Net},\bar{R_{i}})$ $<  Pref(\text{TempCP-Net},  \bar{R_{j}})$. Otherwise, the request is removed from the composition, i.e., $\tau(\bar{R}(N), k) =\bar{R_{j}},\text{if } Pref(\text{TempCP-Net},  \bar{R_{i}}) \geq  Pref(\text{TempCP-Net}, \bar{R_{j}})$. The complexity of finding $C(\bar{R}(N), k)$ is $O(N^{k})$. As there are at most $N$ requests to be considered in a set, we solve the DP from bottom up manner in the following sequence:  $\tau(\bar{R}(N), \;1)$,  $\tau(\bar{R}(N), \;2), \cdots, \tau(\bar{R}(N), N)$.  The final complexity of the DP based solution is $O(N^{O(N)})$.  

    \item \textit{Local sequential optimization approach:} This approach optimizes requests in each time segment. The key reason is that we do not need to consider the whole set of requests during the entire composition period. We only consider a partial set which is applicable in a specific temporal segment. It should reduce the runtime complexity significantly. In Fig \ref{fig:newExample1}, the local optimization could be divided into two segments: optimization with ${A,B}$ in the first year ($OP_{i}$) and optimization with ${A, B, C}$ ($OP_{j}$) in the second year. As the request sets are deterministic, we can perform the optimization sequences in different orders, i.e., $<OP_{i},OP_{j}>$ or $<OP_{j},OP_{i}>$. Local optimizations are dependent on the accepted or rejected requests during previous optimizations in a sequence. For example, if the sequence is $<$1st year, second year$>$ in Fig \ref{fig:newExample1} and we reject request $B$ in the first year, the candidate request set for local optimization is the second year reduces to $<A, C>$.
\end{itemize}

We propose a heuristic based sequential optimization approach for an IaaS composition in \cite{sajibicsoc2016}. To improve the quality of the solution, we develop a reinforcement learning based approach to find the best local service provision policy, i.e., the best selection of requests in the optimal temporal sequence.

\subsection{\textbf{Sequential IaaS Composition using Reinforcement Learning}}

We formulate the long-term IaaS composition problem, i.e., the selection of requests, as a sequential decision process. We begin with a time segment in the TempCP-net and select a request for the composition. The selection of the requests for the next segment depends on the previous selections of requests as accepted overlapping requests are already committed for both the segments.

A sequential decision process is modeled in different approaches such as Multi-Armed Bandit (MAB), Markov Decision Process (MDP), or Partially Observable Markov Decision Process (POMDP) \cite{kaelbling1996reinforcement}. We observe that the state-action-reward situation in the IaaS composition is similar to the MDP. We may start the selection of requests (actions) in a time segment and can compute the local preference ranking of the selected requests by matching the temporal segment in the selected requests with the corresponding temporal \textit{k}-d tree of the given TempCP-net. The local preference ranking may be considered as the reward function. After the selection of requests in a segment, the composition approach may transit to any segment and make new selections. The process may continue until total reward cannot be further maximized (convergence) (as we consider ranking values as the reward, maximizing total rewards refer to minimizing global ranking). MAB or POMDP is also applicable in our context as they are special cases of MDP. For example, MAB is a special case of MDP that has only one state. We select MDP as the general sequential decision process.

\begin{figure}[t!]
	\centering
  	\includegraphics[width=.5\textwidth]{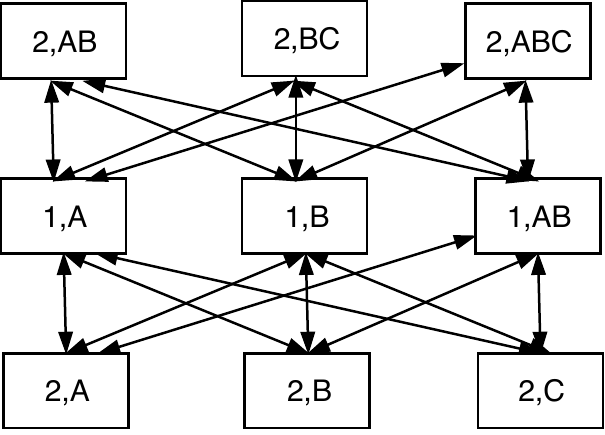}  
    \caption{State-action transitions for sequential composition}
	\label{fig:newExample2}
	\vspace{-4mm}
\end{figure}

Figure \ref{fig:newExample2} shows the all possible state transitions for the request sets in the Figure \ref{fig:newExample1}. The state-action is represented by the pair [interval, Request]. In Fig \ref{fig:newExample2}, [1,A] refers that request A in Fig \ref{fig:newExample2} is selected in the interval 1. We consider bi-directional transition edges. The key reason is it does not specify the transition sequence. For example, if request C is selected in the second interval first, i.e. [2,C], the next transition may happen to any of the [1,A], [1,B] or [1,AB]. [1,AB] represents that both requests A and B are selected in the interval 1.

As the composition environment is dynamic, model-free learning, e.g., reinforcement learning (RL) is usually more applicable than the model-based learning algorithms to implement the MDP. To solve the composition problem using a reinforcement learning (RL) approach, we treat each new request sets as a new environment and learn the optimal selection of requests through multiple interactions with the environment. We focus on the Q-learning approach as a reinforcement learning (RL) approach. Note that other deep learning approaches could be implemented in our context. However, we are not focused on providing a comparative study of machine learning approaches in this paper; instead, we use what we think is a sound approach in our context. Our primary target is to apply an unsupervised machine learning approach in the long-term IaaS composition. In that respect, our primary target is to evaluate the effectiveness of general reinforcement learning, i.e., Q-learning and its proposed variations in our context. In the future work, we will compare the performance of the proposed approach with other deep reinforcement learning approaches.

\subsubsection{\textbf{Q-learning based Approach in IaaS Composition}}

A widely used reinforcement learning is Q-learning. In a Q-learning based method, past interactions in the same environment are utilized to learn the optimal policy \cite{watkins1992q}. The sequences of request selection i.e., \textit{experience} over different intervals can be treated as past interactions in the context of IaaS composition. We define experience as a tuple  $<s, a ,r, \acute{s}>$ where $s$ is the current interval, $a$ is the selected request in $s$, $r$ is the reward for selecting $a$, and $\acute{s}$ is the next interval. We represent the history of interactions as $<s_{0}, a_{0}, r_{1}, s_{1}, a_{1}, r_{2},....>$. We formally define the Q-learning environment in the context of qualitative composition as follows:
 
\begin{itemize}
\item \textit{Environment}: The environment consists of consumers and the provider. The consumer requests are represented into time series groups and the provider's long-term qualitative preferences are represented in TempCP-net. The environment is deterministic, i.e., incoming requests and TempCP-net is given prior to the composition.

\item \textit{State ($s$)}: The tempCP-net is usually consists of several temporal CP-nets for different time intervals or segments. Each time interval or segment is treated as a state. The number of states is finite.  
\item \textit{Action ($a$)}: The selection or rejection of a request is treated as an action in our context. In Figure \ref{fig:newExample1}, the second segment (2nd year) has 3 available requests $\{A,B, \text{ and} \;C\}$. Hence, the possible set of acceptance actions is $\{A,B,C,AB,AC,BC,ABC\}$.  

\item \textit{Policy ($\pi$)}: It is a function that determines which action to perform, i.e., selection of requests in a given state. If $S$ is the set of all states and $A$ is the set of all possible actions, the policy ($\pi$) is represented as $\pi(s): S \longrightarrow A$.

\item \textit{Reward function ($RWD$)}: We match the action, i.e., selected request segments with the corresponding segment in TemCP-net. We consider the local preference ranking as the reward. The reward function is defined based on the equation \ref{eq:ranking} as $RWD(s,a)=Pref(\text{TempCP-net(s)},a)$ at a state $s$ and actions $a$.
 
\item \textit{Value function ($V$)}: $V^{\pi}(s)$ is the state-value function in the sequential decision process. It is the expected cumulative preference ranking starting from state $s$ following a policy $\pi$.
\end{itemize}

We apply the basic Q-learning approach to the IaaS composition. We propose a modified Q-learning approach for the long-term IaaS composition.

\subsubsection{\textbf{IaaS Composition using 2d Q-learning}}

A value function $V^{\pi}(s)$ represents how good is the temporal sequence of a segment for the composer to select requests. The value function depends on the policy by which the composer chooses its actions. Among all possible value-functions, there exists an optimal value function that has a higher value than other state functions:
\vspace{-5mm}

\begin{equation}
    V^{*}(s) = max_{\pi} V^{\pi}(s) \;\;\forall s \in S 
\end{equation}

The optimal policy $\pi^{*}$ corresponds to the optimal value function: \vspace{-3mm}

\begin{equation}
    \pi^{*}= arg\;max_{\pi} V^{\pi}(s) \;\;\forall s \in S 
\end{equation}

A recursive function called $Q$ is usually used in a Q-learning process and represented as $Q^{\pi}(s,a)$ \cite{watkins1992q}. It is used to calculate the cumulative reward using the policy $\pi$. We map $Q^{\pi}(s,a)$ with the expected global preference ranking of choosing the request $a$ in the interval $s$ for the policy $\pi$. The probability of moving from interval $s$ to interval $\acute(s)$ is denoted by $P(\acute{s}| s,a)$ in Equation \ref{eq:q3} where $a$ is the selected requests. The current preference ranking of $a$ in $s$ is denoted by $R(s,a,\acute{s})$ where $\acute{s}$ is the next interval. The future global ranking is denoted by $V^{\pi}(\acute{s})$ in Equation \ref{eq:q3}.


\vspace{-3mm}

\begin{equation}
\label{eq:q3}
Q^{\pi}(s,a) = \sum_{\acute{s}} P(\acute{s}| s,a)(R(s,a,\acute{s})+\gamma V^{\pi}(\acute{s}))
\end{equation}


In a Q-learning process, a table $Q[S,A]$ is maintained where $S$ denotes the set of states and $A$ denotes the set actions \cite{watkins1992q}. $Q[S,A]$ is used to store the current value of $Q^{\pi}(S,A)$. The value of $Q^{\pi}(S,A)$ in the context of long-term IaaS composition is computed using temporal differences. Therefore, we create a table $Q[S,A]$ where $s$ denotes the set of interval or segment and the set of action is denoted by $a$. We set the initial $Q[s,a]$ to 0 for each $(s,a)$. we start the process from an random state ($s$) and executes a random action ($a$) for a reward $r$. The next interval is also selected randomly. We use $\epsilon$-greedy policy to restrict the randomness over the time. The idea is the composer should explore the state-action sequences randomly, in the beginning, to find better future discounted preference rankings, later the randomness should be reduced. Here, $\epsilon$ is defined as the probability of exploration. The exploration is equivalent to picking a random action in action space. If $\epsilon$ = 1, the composer will always explore, and never act greedily concerning the action-value function. Therefore, $\epsilon < 1$ in practice, so that there is a good balance between exploration and exploitation. The higher value of alpha usually assigns higher weights to the current estimate than the previous estimate. The learning process terminates when there are no further updates on Q-values. It is also known as convergence values. Once the Q-learning reaches the convergence state, the optimal policy is found using $Q[S,A]$ in Equation \ref{eq:q5}. At each segment the action with highest reward is selected. An example of 2d $Q[S,A]$ is shown in Figure \ref{fig:f7} where the number of segments is 5 and the number of actions is 10. The best action in segment 5 is $A10$ or $A5$.


\begin{equation}
\label{eq:q4}
Q[s,a] = (1-\alpha)Q[s,a] + \alpha (r+\gamma max_{\acute{a}}Q[\acute{s},\acute{a}]) 
\end{equation} 


\begin{equation}
\label{eq:q5}
\pi^{*}(s) = argmax_{a}Q[s,a] \;|\; \forall a \in A(s)
\end{equation}

\subsubsection{\textbf{IaaS Composition using 3d Q-learning}}

\begin{figure}[t!]
  	\centering
  	\includegraphics[width=0.8\textwidth]{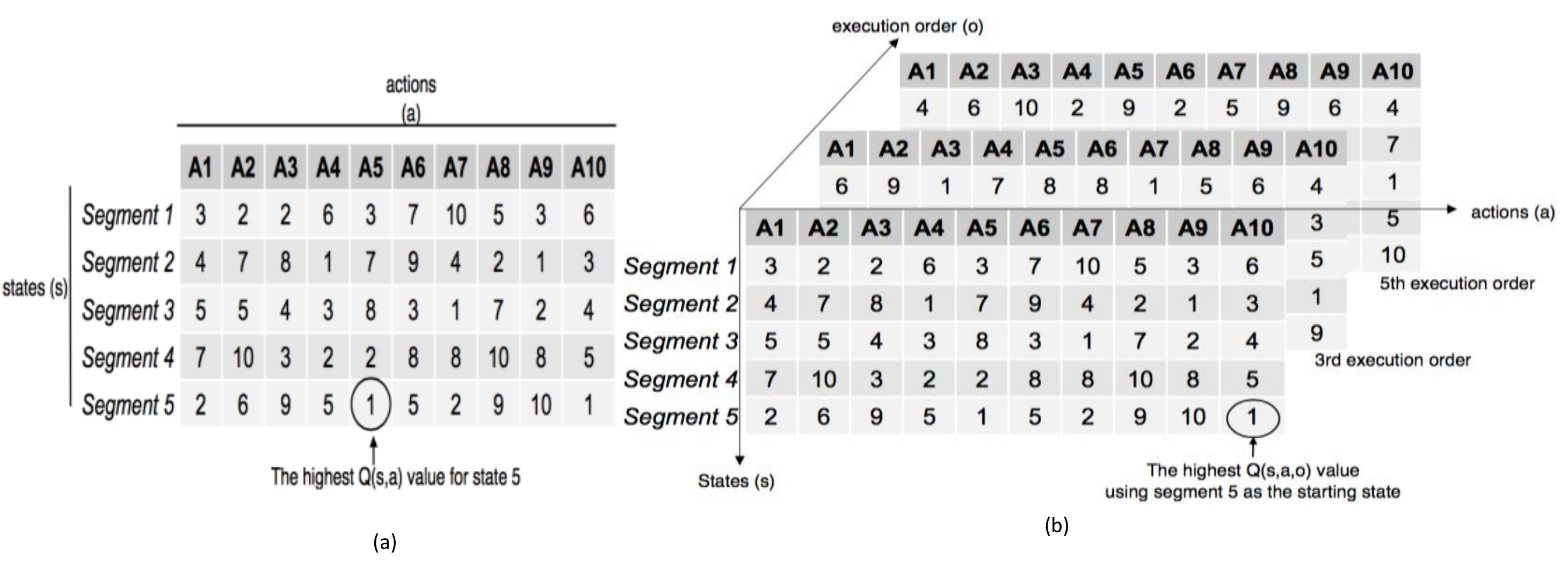}   
    \caption{(a) Q-values in a 2d $Q[S,A]$ (b) Q-values in a 3d $Q[S,A, O]$}
    \vspace{-3mm}
	\label{fig:f7}
\end{figure}

The 2d Q-learning has no start and terminal states as it accepts only model-free state transitions \cite{shani2005mdp}. The long-term effect of sequential order is indicated by $Q[S,A]$. However, it is not possible to keep track of the execution order in the 2d Q-learning process. For example, there are several possible state transitions can take places in Figure \ref{fig:newExample2} such as $\{[1,A]\rightarrow [2,B]\}$ and  $\{[2,B]\rightarrow [1,A]\}$. Here, $Q[1,A]$ represents the expected global preference rankings of selecting action $A$ in the first year irrespective of the sequence of selection (first or second). A similar explanation is also applied to $Q[1,A]$. Besides, the existing Q-learning approaches for the composition allow multiple execution order. Such order in Figure \ref{fig:newExample2} is  $\{[1,A] \rightarrow [2,BC] \rightarrow [1,B]\}$. Note, the selection of request $A$ and $B$ in the first year is taken at two different positions in a sequence.


We introduce a three-dimensional Q-learning in \cite{mistry2018long} using a 3d table $Q[S,A,O]$ to store the Q values. $O$ represents the set of execution orders. Therefore, a particular state-action pair $(s,a)$ may have different values depending on the execution order $o$. If a set of requests is selected from the first segment at the first step of the composition, it may have a different preference ranking than if it is selected at the last step of the composition. Figure \ref{fig:f7} illustrates an example of 3d $Q[S,A,O]$. Here, the ranking of $A10$ is 1 in segment 5 when it is performed as the starting state. However, the preference ranking of $A10$ changes into 9 when it is performed in the third step. An extension of Equation \ref{eq:q4} is shown in Equation \ref{eq:q6} for a 3d Q-learning process. The $\acute{o}$ denotes the next execution order after $o$ and $\alpha$ denotes the learning rate. The 3d Q-learning selects the start state ($s$), perform an action ($a$) with reward $r$ arbitrarily. From the start state, it observes all the possible states in different orders.

\vspace{-4mm}

\begin{align}
\label{eq:q6}
Q[s,a,o] = &(1-\alpha)Q[s,a,o] \\ \notag
&+ \alpha (r+\gamma max_{(\acute{a},\acute{o})}Q[\acute{s},\acute{a},\acute{o}]) 
\end{align}

\subsubsection{\textbf{On-policy based 3d Q-learning}}
\begin{figure}
  	\centering
  	\includegraphics[width= .8\textwidth]{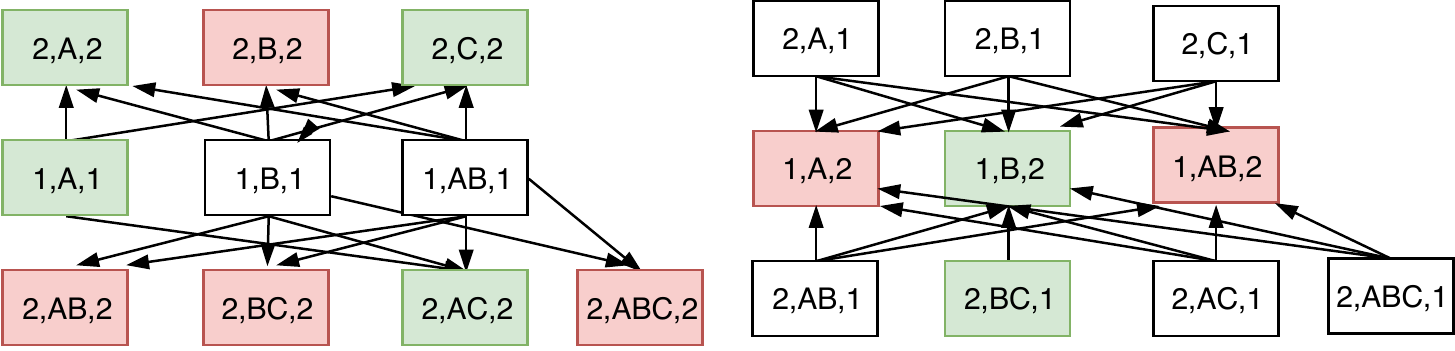}   
    \caption{On-policy state-action transitions in 3d Q-learning (Green colored are allowed from [1,A,1] and [1,B,2]}
    \vspace{-4mm}
	\label{fig:newexample3}
\end{figure}

The 3d Q-learning increases the number of explorations in the learning process than the 2d Q-learning. The $n\times m$ Q-matrix from two dimensions is extended to $n\times m \times p$ Q-matrix. The number of states is denoted by $n$, the number of actions is denoted by $m$ and $p$ is the number of segments. As the exploration space increases, 3d Q-learning requires more time to learn compared to the 2d Q-learning. The 3d Q-learning based composition approach is considered as off-policy, as the composer has no restrictions over exploration. We transform the off-policy approach into an on-policy learning approach by intelligent reduction of state transitions as follows:

\begin{itemize}
    \item \textit{State sequence policy}: Once a request is rejected, it is not considered anymore in following local optimization. Therefore, we do not need to consider the same states multiple times. We introduce state sequence policy where each state is visited only once in a policy $\pi$. 
    \item \textit{Removing redundant state transitions}: The rejected request in a local segment may appear in another segment if the request overlaps the two segments. All the state-action pairs that contain the rejected request should not be considered as the next state transitions. For example, if we accept request $A$ in the first year, request $B$ is rejected in Figure \ref{fig:newExample1}. However, request $B$ overlaps into the second year. Hence it should be removed from the next candidate transitions. Figure \ref{fig:newexample3} depicts the on-policy state transitions when the request $A$ in the first year is selected at first in the sequence.        
\end{itemize}

We represent the on-policy 3d Q-learning process for IaaS composition in Algorithm \ref{alg:qlearning}. Algorithm \ref{alg:qlearning} runs multiple episodes or round to perform the 3d Q-learning process. The algorithm uses Equation \ref{eq:q6} to update the Q-values in each episode. A $\epsilon$-greedy policy is used by algorithm \ref{alg:qlearning} where the optimal action is selected for execution. The optimal action is denoted by $arg\;max_{(a,o)}Q(s, a,o))$ where the probability of the optimal action is $(1 - \epsilon)$. The greedy policy optimizes $Q(s, a,o)$ continuously. It incorporates the unique state sequence policy. When a request is rejected, we remove it from the candidate action sets. The learning process continues in a loop up to a maximum number $k$ or Q-values converge to the optimal value. Once the Q-values reach convergence state, we use Equation \ref{eq:q7} to compute the optimal policy which is similar to 2d Q-learning process \cite{wang2010adaptive}.  

\vspace{-4mm}
\begin{align}
\label{eq:q7}
\pi^{*}(s) = &argmax_{(a,o)}Q[s,a,o] \\ \notag
&\text{where } \forall a \in A(s) \text{ and } o \in [1, |A(s)|]
\end{align}

\begin{algorithm}
\fontsize{8pt}{8pt}\selectfont
    \caption{The on-policy 3d Q-learning process to compose IaaS requests}
    \label{alg:qlearning}
    \begin{algorithmic}[1]
    \STATE Initialize Q(s,a,o) to 0
    \FOR {each episode to $K$}
    \STATE $s \gets s_{0}$
    \STATE execution order, $o\gets 1$
    	\WHILE{$o \neq$ total number of segments}
    	 	\STATE Choose action $a$ from $s$ in $o$ using $\epsilon$-greedy policy.
    	 	\STATE  Execute $a$, observe reward $r$ and next state $\acute{s}$
    	 	\STATE $Q[s,a,o] \gets (1-\alpha)Q[s,a,o] + \alpha (r+\gamma max_{(\acute{a},\acute{o})}Q[\acute{s},\acute{a},\acute{o}])$
    	 	\STATE create candidate $[s,a]$ based on redundant state transitions
    	 	\STATE $s \gets \acute{s}$ using $\epsilon$-greedy policy. 
    	 	\STATE  $o \gets o+1$ 
		\ENDWHILE
    \ENDFOR
    \end{algorithmic}
    
\end{algorithm}

\section{Long-term Qualitative Composition with Previous Learning Experiences}

We aim to utilize the knowledge of composing past requests to compose new incoming requests efficiently. In this section, we describe the proposed long-term qualitative composition approach that leverages past learning experience. We analyze different types of requests patterns and illustrate how to identify similar request sets. Once we identify a similar requests set from the past incoming requests, we reuse previously learned policy to compose new incoming requests. 

\begin{figure}
    \centering
      \includegraphics[width=.9\textwidth]{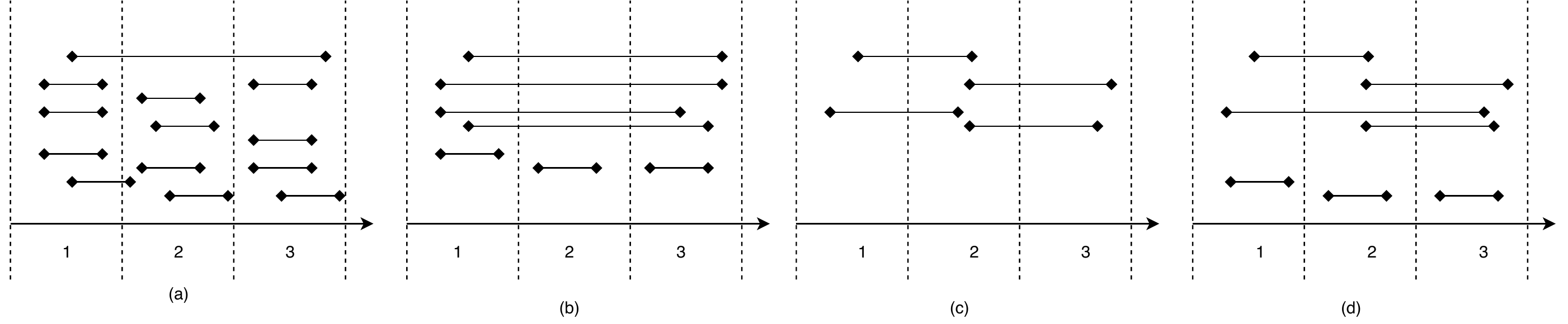}   
    \caption{Different requests patterns: a) almost sparse pattern, b) almost dense pattern, c) chain, (d) mixed patterns \cite{sajibicsoc2016}}
    \vspace{-4mm}
    \label{fig:pattern}
\end{figure}

The optimal sequence of state-action may vary depending on the distribution of the requests over the time and their rankings. For example, Figure \ref{fig:pattern} shows four types of request patterns, i.e., almost sparse pattern, almost dense pattern, chain pattern, and mixed pattern.

\begin{itemize}
    \item \textit{Almost Sparse Pattern}: Figure \ref{fig:pattern}(a) shows a set of \textit{almost sparse  request pattern}. Most requests are short and disjoint, i.e., do not overlap between two intervals. The composition may be performed in parallel by taking the short-term requests.
    \item\textit{Almost dense Pattern}: Figure \ref{fig:pattern}(b) shows a set of requests where the requests are mostly \textit{overlapped} between intervals. An overlapping request spans for the next interval. Most requests are accepted in the first step of the selection.
    \item\textit{Chain Pattern}: Figure \ref{fig:pattern}(c) shows a chain pattern where the requests are mostly short-term and overlapped between intervals. The requests are also evenly distributed between the intervals in a chain pattern.
    \item \textit{Mixed Pattern}: Figure \ref{fig:pattern}(d) shows a mixed pattern where both long-term and short-term requests are overlapped and evenly distributed.
\end{itemize}

\begin{figure}
    \centering
      \includegraphics[width=.6\textwidth]{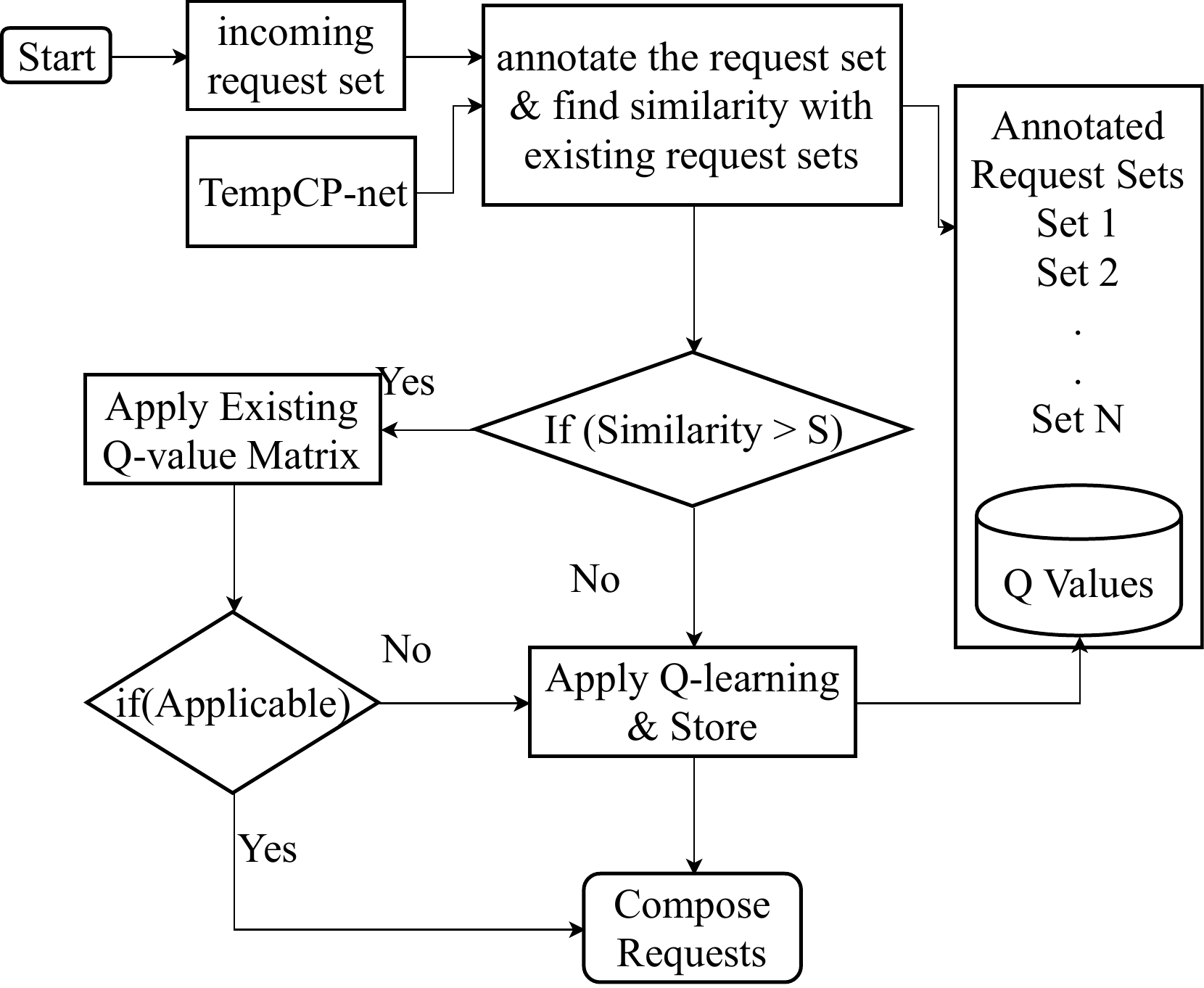}   
    \caption{A qualitative composition framework using policy library from Q-learning}
      \vspace{-4mm}
    \label{fig:flowchart}
\end{figure}

Applying the Q-learning method each time when a new set of requests arrives can be expensive. Instead of learning a new set of incoming request every time, we apply the experience from the previously learned request sets. We propose a qualitative composition framework as shown in Figure \ref{fig:flowchart}. The proposed framework takes a set of incoming requests and the TempCP-net of the provider. First, the given request set is annotated with its global preference ranking and overlapping ratio and matched with the existing sets of requests. Initially, there is no existing request set. A Q-learning algorithm is applied to the request set using the TempCP-net. The output of the Q-learning algorithm is a matrix called Q-value. The learned Q-value matrix is stored with the corresponding request set for future. 

Each time a new set of requests arrives the proposed framework find the similarity with existing request sets. The similarity is measured through a hierarchical clustering method called \textit{agglomerative clustering method} \cite{Growendt2017,fernandez2008solving,bouguettaya2015efficient}. For each set of requests, we apply the\textit{ agglomerative clustering method} to build a corresponding clustering tree. To measure the similarity between two request sets, the correlation coefficient of their corresponding clustering trees is computed. If the similarity is greater than a predefined variable $S$, then the Q-value matrix of the corresponding matched request set is applied to compose the new request set. We set the value of $S$ based on the trial and error in the experiment. The existing Q-value matrix may not be fully applicable to a new set of requests. In such a case, the proposed framework applies the Q-value matrix partially (a policy reuse approach) and learns the rest of the sequence.

\label{sec:dq}


In our previous work, we calculate similarity between different types of requests concerning the statistical distribution of their resources attributes such as normal, left-skewed and right-skewed distributions of CPU, memory, and network bandwidth \cite{mistry2018long}. The proposed approach however is unable to capture intrinsic characteristics of different request patterns such as temporal distribution or global ranking. Therefore, it may not correctly utilize the historical information of the previous consumers' request sets.

\subsection{\textbf{Clustering Methods to Find Similar Request Sets}}

We use clustering techniques to compute the similarity between a new request set and past request sets. Clustering is a well-known data analytic techniques to capture the intrinsic features of a set of data and group them on different sets based on similarity. Clustering is suitable where manual tagging of data is expensive. Moreover, the prior knowledge for manual tagging may not be available or insufficient. In such a case, clustering is a preferred option over supervised learning approaches such as classification and regression \cite{Growendt2017,fernandez2008solving,bouguettaya2015efficient}. There are many clustering techniques in the existing literature. We focus on partitional and hierarchical clustering approaches in the IaaS composition. 

\subsubsection{\textbf{Partitional Clustering}}

Partitional clustering methods produces a flat partition of that optimizes an objective function. The objective function needs to be predefined. The most well-known partitional algorithm is \textit{K-means} clustering. The main steps of K-means clustering of a request set are as follows:
\begin{enumerate}
    \item Create randomly few centroid points for the requests. 
    \item Assign each request to the closest centroid. 
    \item Calculate the central point of the newly created clusters and update the centroid accordingly
    \item Repeat the previous last two steps until there is an object left to resign to another cluster. 
\end{enumerate}

The computational complexity of the K-means clustering is $O(NK)$ where $N$ is the number of requests and $K$ is the number of the clusters. The K-means clustering is efficient concerning computational clustering as it requires linear computational time. However, the performance of K-means depends on how the value of $K$ is chosen. It is difficult to determine the optimal value of $K$ when prior knowledge is inadequate or absent.

\subsubsection{\textbf{Hierarchical Clustering}} 


The hierarchical clustering method is a mainstream clustering method because it can be applied to the most types of data. Although hierarchical clustering method has a higher complexity compare to the K-means, it does not need predefined parameters. Therefore, hierarchical clustering are more suitable for handling real-world data. There are two main approaches for hierarchical clustering bottom up and top down. The bottom-up approach aggregate individual data points to the most high-level cluster. The complexity is usually $O(N^2)$ for the bottom-up approach. However, it may go up to $O(N^2logN)$. The complexity of the top-down approach is $O(2^N)$. The top-down approach is usually more expensive than a bottom-up approach. The bottom-up approach is generally known as \textit{agglomerative hierarchical clustering} \cite{murtagh2012algorithms,bouguettaya1996line} . 

\subsubsection{\textbf{Agglomerative Clustering based Similarity Measure}} 

We use an agglomerative clustering based approach to reuse the existing policies for a new incoming request set based on the history of the past request sets. The clustering approach is applied to a set of requests to construct a clustering tree. The clustering tree captures the intrinsic features of the requests and group the requests based on their similarities. When a set of requests arrives, we build a clustering tree and compare it with the existing clustering trees to find the most similar clustering tree using the correlation coefficient.

We annotate each request in a set of requests with its global rank and overlapping ratio to construct a clustering tree to capture the temporal aspects and global ranking of a request set. The global rank is computed by the equation \ref{eq:ranking}. The overlapping ratio of a request is the ratio of the number of operation interval and the number of total intervals of the composition. The overlapping ratio of a request $R_i$ is computed by the following equation: 

\begin{equation}
    O(R_i) = \frac{\text{Number of intervals of } R_i}{\text{Total Number of Intervals}}
\end{equation}

\begin{figure}[t!]
    \centering
    \includegraphics[width=.7\textwidth]{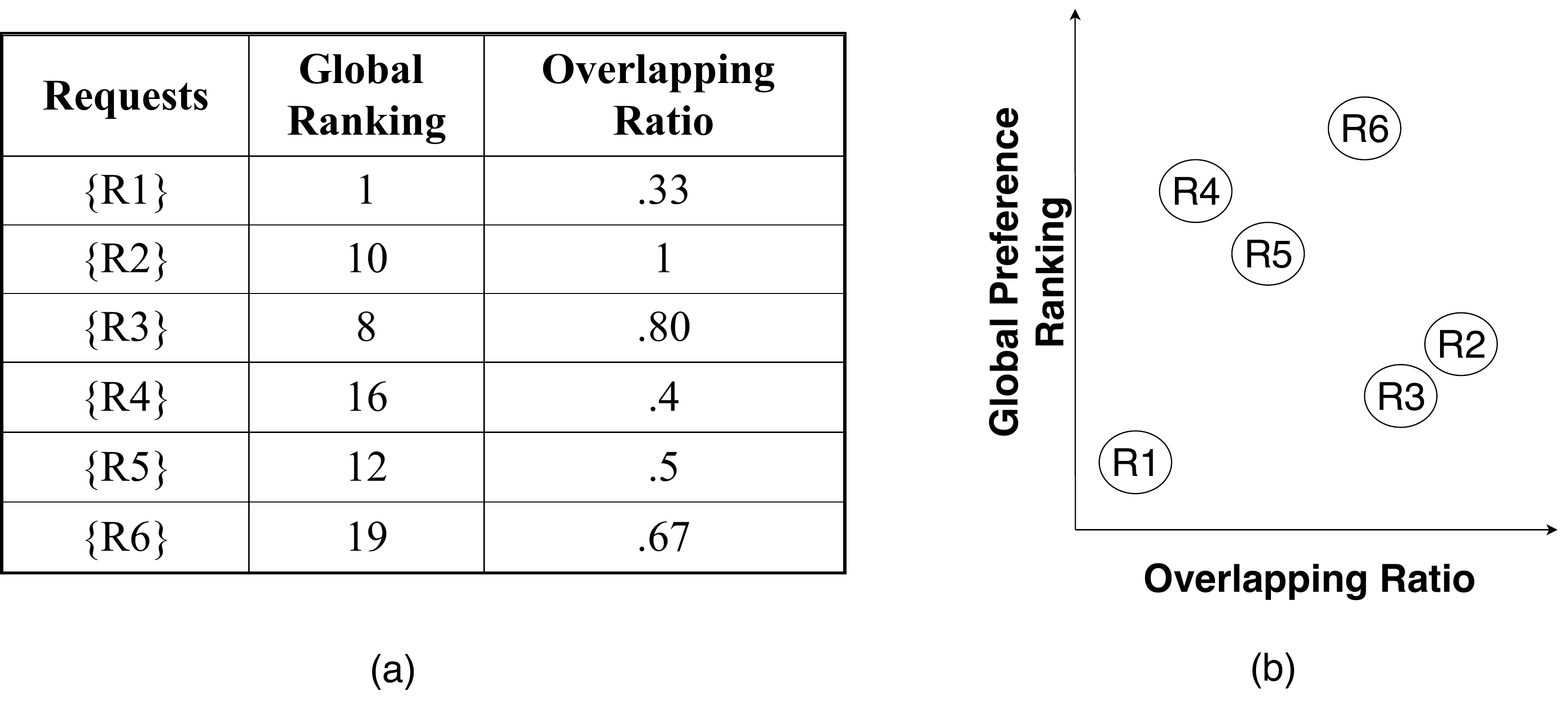}
    \caption{Annotation of a request set using global preference ranking and overlapping ratio (a) annotation table (b) annotation plot}
    \label{fig:plot}
    \vspace{-2mm}
\end{figure}

The overlapping ratio for the request $R3$ in the request set $A$ (Figure \ref{fig:req}) is $2/3=.667$. Let us assume a request set is annotated with the global preference ranking and the overlapping ratio of each request as shown in Figure \ref{fig:plot}. Now, we construct the clustering tree based on the following steps:

\begin{enumerate}

    \item Each request is considered as a cluster. If there are $N$ requests in a set of requests, then the number of clusters is $N$.
    \item A $N X N$ distance matrix is constructed based on the \textit{Euclidean Distance} (Equation \ref{eqn:euclid}) of each pair of requests according to their global preference ranking $GPR$ and overlapping ratio $OR$. 
    \item The closest pair of clusters is selected and merge them into a single cluster.
    \item The distances between the new cluster and each of the old clusters are computed.
    \item Steps 3 and 4 are repeated until there exists only a single cluster. 
\end{enumerate}

\vspace{-3mm}

\begin{equation}
    E.D = \sqrt{\sum{GPR(R1-R2)}^2+OR(R1-R2)^2}
    \label{eqn:euclid}
\end{equation}

We can perform step 4 in different ways based on different hierarchical clustering approaches \cite{murtagh2012algorithms,fernandez2008solving,Growendt2017,bouguettaya1996line}. We consider different conventional approaches to measure the distance between two clusters which are as follows:

\begin{enumerate}
    \item SLINK: SLINK stands for single linkage clustering method. In this method, two clusters are joined based on the distance of their nearest pair of elements. Only one member of each cluster is considered to compute the distance \cite{murtagh2012algorithms}. It is also known as \textit{nearest neighbour}(NN) clustering method.  
    \item CLINK: CLINK is short for complete linkage clustering method. Two clusters are joined based on the distance of the farthest pair of elements \cite{Growendt2017}. It is also known as \textit{farthest neighbour}(FN) clustering method.  
    \item UPGMA: This unweighted pair-group method using arithmetic average is also known as the average linkage clustering method \cite{fernandez2008solving}. We compute the distance between two clusters based on the average distance of each pair of elements from two clusters.
\end{enumerate}

\begin{figure}[t!]
    \centering
    \includegraphics[width=0.9\textwidth]{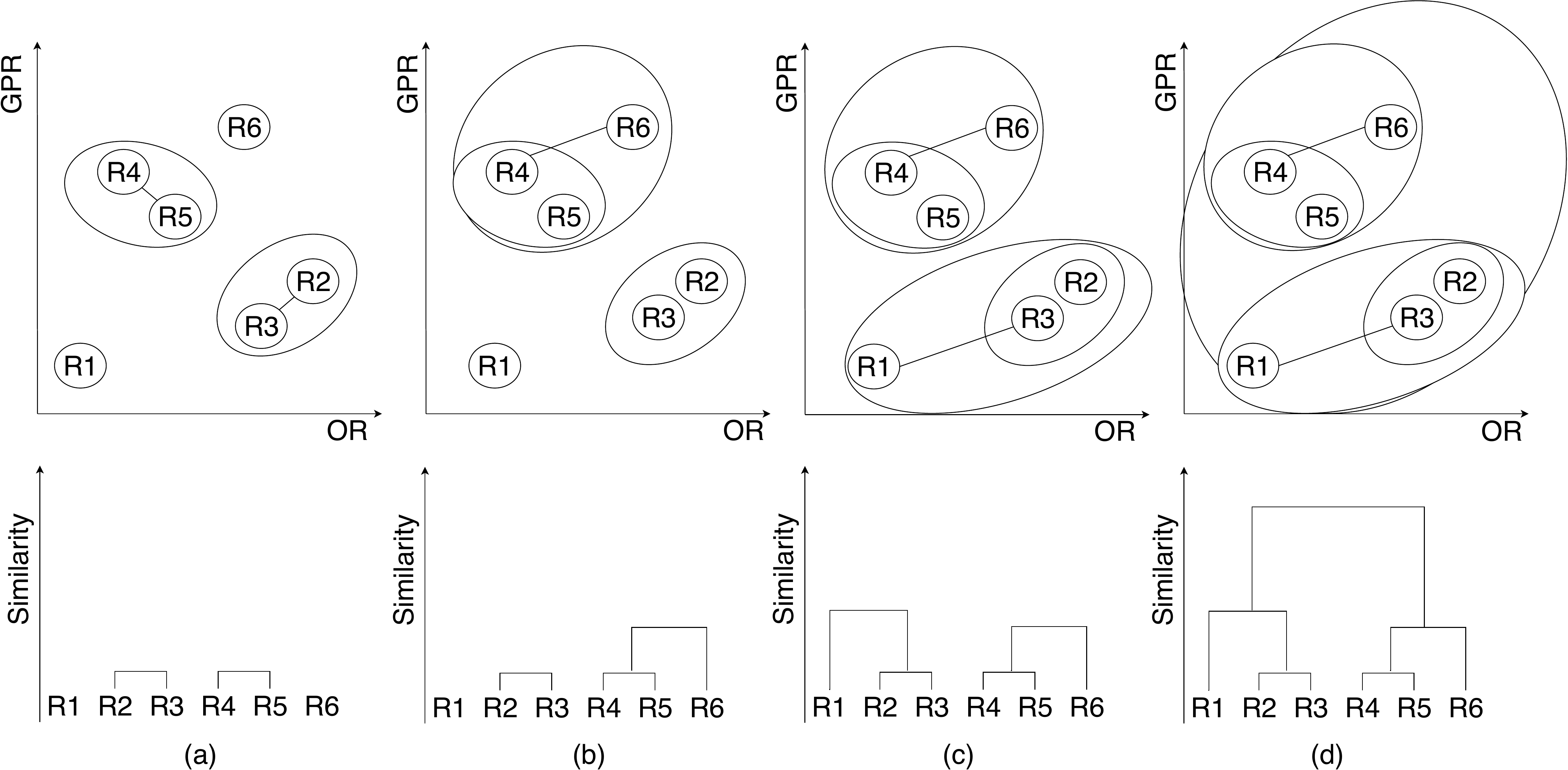}
    \caption{Hierarchical clustering steps}
    \label{fig:clusters}
     \vspace{-3mm}
\end{figure}

Any of the above hierarchical clustering approaches, i.e., SLINK, CLINK, UPGMA could be used for the IaaS composition of new requests. We use the SLINK or nearest neighbor approach as it is a widely used clustering approach and effective in time-series data clustering \cite{berkhin2006survey}. \textit{Note that, finding the optimal clustering approach for IaaS composition is out of the focus of this paper}.

A clustering construction process is shown in Figure \ref{fig:clusters}. Initially, we perform the annotation as shown in Figure \ref{fig:plot}. In Figure \ref{fig:clusters}(a), \{R4\} and \{R5\} are the nearest to each other. \{R4\} and \{R5\} are put in the same cluster. Similarly, \{R2\} and \{R3\} are the nearest to each other and joined in the same cluster. In Figure \ref{fig:clusters}(b), \{R6\} is the nearest to the \{R4\}. Therefore, \{R6\} is joined with \{\{R4\},\{R5\}\}. In the next step, \{R1\} is the nearest to \{R3\}. \{R1\} is joined with \{\{R2\},\{R3\}\}. In Figure \ref{fig:clusters}(c), there are only two clusters which are \{\{R4\},\{R5\},\{R6\}\} and \{\{R1\},\{R2\},\{R3\}\}. These two clusters are joined based on \{\{R5\} and \{R2\} who are the nearest request to each other from two clusters. Finally, there is only one cluster left.

Once we build clustering trees for different sets of requests, we need to compute the coefficient of correlation between different clustering trees. We use \textit{Cophenetic correlation coefficient} to compare two clustering trees \cite{sokal1962comparison}. A cophenetic correlation coefficient determines how well a clustering tree preserves the pairwise distance between the original requests before clustering. \textit{The cophenetic distance between two requests in a clustering tree is the height of the clustering tree where the two branches that include the two requests merge into a single branch.} We compute the cophenetic correlation coefficient for each clustering tree and use them to measure similarities between two clustering trees. Given a set of requests R, its corresponding clustering tree T, R(i,j) is the Euclidean distance between $i$th and $j$th requests, and T(i,j) is the cophenetic distance where two requests merged first time in the clustering tree, we compute the cophenetic correlation coefficient using the following equation \cite{sokal1962comparison}: 
\begin{equation}
    c = \frac{\sum_{i<j}(R(i,j)-\bar{R})(T(i,j)-\bar{T})}{\sqrt{[\sum_{i<j}(R(i,j)-\bar{R})^2][\sum_{i<j}(T(i,j)-\bar{T})^2]}}
    \label{eqn:cop}
\end{equation}

\subsection{\textbf{Q-learned Policy Reuse in IaaS Composition}}
Let us assume we are given a library of Q-matrices where the corresponding incoming requests set has a cophenetic distance to a new set of requests, $c$ below a certain threshold $ThC$. Our target is to reuse the Q-matrices to find the optimal composition of the new set of requests. We reuse existing policy library in two ways: a) greedy policy reuse approach, b) policy reuse in on-policy 3d Q-learning approach:

\subsubsection{\textbf{Greedy Policy Reuse Approach}}

The best policy reuse approach should not involve Q-learning approach for a new set of requests at all. However, it is challenging to apply an existing policy directly to a new set of requests due to the new request sets, e.g., new states and actions in the environment. However, the number of temporal segments remain the same as the provider's long-term TempCP-net does not change. As a greedy approach we can partially reuse the policy, i.e., only the temporal sequences from the previous history could be applied to perform local optimization. Note that, the greedy policy reuse approach does not involve any reinforcement learning to compose a new set of requests. The steps for greedy policy reuse approach are described below:

\begin{enumerate}
    \item Find the similar Q matrices where cophenetic distance $c$ is below a certain threshold $ThC$ using the Agglomerative Clustering approach.
    \item Select the Q matrix that is learned for the request set of min($c$).
    \item Generate the optimal policy ($\pi^{*}$) from the Q matrix.
    \item Let us assume  $D$ is the sequence of state transitions in ($\pi^{*}$).
    \item Following the sequence $D$, perform local optimization using Dynamic programming in each segment. Accept the requests from the output of the optimization. Rejected requests are removed from the candidate list for the next available sequence.
    \item Return the set of accepted requests when local optimization is performed in all segments in $D$. 
\end{enumerate}

\subsubsection{\textbf{Policy Reuse in the On-policy 3d Q-learning Approach}}

We omit the reuse of learned actions, i.e., selection of requests in the greedy approach. Here, we focus on policy reuse in the exploration process in the proposed Q-learning approach. We adopt the policy reuse approach for two-dimensional Q-learning approach \cite{fernandez2006probabilistic} for the proposed on-policy 3d Q-learning approach. The key idea in policy reuse is to integrate the past policy as a probabilistic bias in the exploration strategy of the new learning process for composition. 

We define $\mu$ as the policy exploitation probability that defines the exploitation of past policy, the new policy or a random exploration. Hence, the $\epsilon$-greedy state transition is replaced with $\mu$-$\epsilon$-greedy state transition as follows:
\begin{enumerate}
    \item Follow the past policy $\pi_{past}(s)$ to select the next state-action with the probability $\mu$
    \item Exploit the current policy $\pi_{past}(s)$ to select the next state-action with the probability $(1-\mu)\epsilon$.
    \item Select the next state-action randomly with the probability $(1-\mu)(1-\epsilon)$.
\end{enumerate}

One issue in the policy reuse is to find the corresponding actions as the partial request, i.e., request in an interval may not be an exact match with the new request. We perform time-series similarity measure \cite{sajibtsc2015} between the request sets. If the similarity index is higher than a threshold $(TMS)$, we consider a one to one mapping between the actions and apply the past policy to the new request sets. If we have a library of $N$ Q matrices, we modify the on-policy 3d Q-learning approach with policy reuse strategy in the Algorithm \ref{alg:qlearningp}.

\begin{algorithm}[t!]
\fontsize{8pt}{8pt}\selectfont
    \caption{The on-policy 3d Q-learning process with policy reuse}
    \label{alg:qlearningp}
    \begin{algorithmic}[1]
    \REQUIRE $L$ is library of $N$ Q matrices shorted on cophenetic distance $c$
    \REQUIRE $H$ is library of $N$ past optimal policy  $\pi_{past}(s)$ from the Q matrices.
    \REQUIRE $E$ is the extra episodes for random exploration
    \REQUIRE the policy exploitation probability $\mu$
    \STATE Initialize Q(s,a,o) to 0
    \FOR {each episode to $N+E$}
    \STATE $s \gets s_{0}$
    \STATE execution order, $o\gets 1$
        \WHILE{$o \neq$ total number of segments}
             \STATE Choose action $a$ from $s$ in $o$ using $\mu$-$\epsilon$-greedy policy.
             \STATE  Execute $a$, observe reward $r$ and next state $\acute{s}$
             \STATE $Q[s,a,o] \gets (1-\alpha)Q[s,a,o] + \alpha (r+\gamma max_{(\acute{a},\acute{o})}Q[\acute{s},\acute{a},\acute{o}])$
             \STATE create candidate $[s,a]$ based on redundant state transitions
             \STATE $s \gets \acute{s}$ using $\mu$-$\epsilon$-greedy policy. 
             \STATE  $o \gets o+1$ 
        \ENDWHILE
    \ENDFOR
    \end{algorithmic}
        
\end{algorithm} 

\section{Experiments}
\label{exper}

In this section, we discuss our experiments to evaluate the proposed approach. First, we discuss different techniques to evaluate the proposed approach. Next, we discuss the experiment setup with real world datasets. Finally, the experiment results and their evaluation are discuss.

We conduct a set of experiments to evaluate the proposed long-term qualitative composition framework in terms of accuracy and runtime efficiency. First, we analyze the efficiency of the proposed long-term qualitative composition framework for a new set of requests (without history). We compare the proposed approaches with four state-of-the-art techniques: a) Global Dynamic Programming \cite{zou2012qos}, b) 2-d Q-learning \cite{moustafa2013multi}, c) Heuristics based optimization \cite{sajibicsoc2016}, and d) On-policy SARSA learning approach \cite{wang2017integrating}. 


\begin{itemize}
    \item \textit{Global Dynamic Programming}: A QoS-aware dynamic service composition approach is proposed in \cite{zou2012qos} that utilizes the concept of global dynamic programming (DP). The global DP approach finds the all possible solution through developing sub-optimal solutions considering the provider's economic model. The global DP approach is able to find the best solution if exists. However, it poses scalability issue during runtime due to comparisons among a large number of candidates. 
    
    \item \textit{2-D Q-learning}: A multi-objective service composition approach is proposed in \cite{moustafa2013multi} that leverages reinforcement learning technique to deal with conflicting objectives and various QoS constraints. The proposed approach uses reinforcement learning to deal with the uncertainty characteristic inherent in open and decentralized environments.  
    
    \item \textit{Heuristic based optimization}: A heuristic based approach is proposed in \cite{sajibicsoc2016} that utilizes sequential optimization. The proposed approach considers several sequences of local optimizations, i.e., left to right, right to left and random. During each iteration, higher utility requests (i.e., trade-off ratio of higher rankings with smaller request length) are selected over higher utility requests. Finally, long-term requests are accepted through collaborative decisions of local optimizations, i.e., based on the global utility score.
    \item \textit{On-policy SARSA learning approach}: A modified SARSA (State-Action-Reward-State-Action) algorithm is proposed as the on-policy reinforcement learning approach for adaptive service composition \cite{wang2017integrating}. The difference between SARSA and Q-learning is that SARSA selects requests (action) following the same current policy and updates its Q-values. However, Q-learning applies the greedy policy, i.e.,  updates its Q-values based on the maximum preference rankings of available actions.
\end{itemize}

The efficiency of the proposed long-term qualitative composition framework is analyzed for a given set of historical request sets. \textit{We consider the outputs of the brute-force approach as the baseline or validation set}. In the brute-force approach, the minimum preference ranking of all the combinations of requests over the total composition period are computed using the global preference ranking Equation \ref{eq:ranking} and pairwise comparisons. As the brute-force approach has exponential runtime, it is not applicable in real world applications. We compare the performance of proposed the greedy policy reuse and the 3d Q-learning with policy reuse approach when the Agglomerative clustering is used for similarity measure. We compare the proposed approach with a Kolmogorov-Smirnov (K-S) test based composition approach \cite{mistry2018long} which is a well-known technique to find correlations between incoming requests and the set of historical requests based on statistical distribution. The historical learned Q-values are used as the initial values (rather than 0) in the K-S test based Q-learning approach \cite{mistry2018long}.


\subsection{\textbf{Simulation Setup with Real World Datasets}}
\label{ex:sim}

Long-term consumer requests are generated from publicly available datasets. Google cluster trace \cite{clusterdata} is used to generate functional requirements of the consumer requests. Google cluster traces provide CPU and Memory workloads of 70 jobs. It does not include the QoS requirements of IaaS consumers. WS-dream dataset \cite{JiangICWS2012} is a real-world cloud QoS performance datasets consist response time and throughput information of one hundred cloud services. As we consider both the functional and non-functional (QoS) attributes in IaaS requests, we generate 70 long-term IaaS requests by creating a random one-to-one mapping between Google cluster and WS-dream dataset. Note that WS-dream dataset does not include IaaS requirements on two QoS attributes, i.e., availability and price which are important for the proposed economic model-based IaaS composition framework. Hence, we randomly generated availability and price values for each IaaS requests.


\begin{table}[t!]
\centering
\caption{Requests distribution description}
\label{tab:request}
\scalebox{1}{
\begin{tabular}{|c|c|c|c|}
\hline
\multirow{2}{*}{\begin{tabular}[c]{@{}c@{}}Distribution \\ type\end{tabular}} & \multicolumn{3}{c|}{\begin{tabular}[c]{@{}c@{}}Request length \\ (in months)\end{tabular}} \\ \cline{2-4} 
 & 1-3 & 4-8 & 9-12 \\ \hline
Normal & 20\% & 60\% & 20\% \\ \hline
Right-skewed & 20\% & 20\% & 60\% \\ \hline
Left-skewed & 60\% & 20\% & 20\% \\ \hline
Random & random & random & random \\ \hline
\end{tabular}
}
\end{table}

Finding actual IaaS provider's business strategy is difficult in the real world. Therefore, Ten tempCP-nets are synthesized where each CP-net has twelve intervals. We consider six attributes namely CPU, memory, response time, availability, price, throughput, and request length. A decision variable is used to represent request length. We generate the dependency randomly to build different types of business strategy. We define ten levels of semantic to generate semantic preference tables. The preference statements for the CPT of each attribute is created randomly. The requests are categorized into four different distributions which are normal, right-skewed, left-skewed, and random. When most of the consumers request the median length of services, we consider the distribution as a normal distribution. Similarly, temporal higher overlapping requirements generate right-skewed distribution and temporal lower overlapping requirements generates a left-skewed distribution \cite{mistry2018long}. For each category of the distribution, we generate five request sets where each set has ten to thirty requests. Table \ref{tab:request} shows the description of each distribution. To create a normal distribution of requests, sixty percent of the total requests are segmented into 4 to 8 month intervals. Rest of the requests are segmented into 1 to 3 month interval and 9 to 12 month intervals. We create other distributions in a similar manner. The starting time of the requests in distribution is generated using poison distribution.

\subsection{\textbf{Efficiency of the Proposed Framework for a New Set of Requests}}

\subsubsection{\textbf{Evaluating Accuracy}}
\label{ex:s1}

In the first experiment, we do not consider the history of Q-learned values. Each request set is composed individually against 10 different TempCP-nets. Initially, moderate learning rate ($\alpha =0.5$) is used for both off-policy and on-policy Q-learning based composition approaches. The outputs are normalized, averaged and compared against the brute force approach. We define the accuracy of a composition as the normalized preference ($NP$):

\begin{equation*}
    NP = \frac{\text{ranking from a composition approach}}{\text{ranking from the brute-force approach}} 
    \label{eqn:cop1}
\end{equation*}

  
  


\begin{figure}[t!]
    \vspace{-8mm}
      \subfloat[]{\includegraphics[width=0.50\textwidth, height=0.5\textwidth]{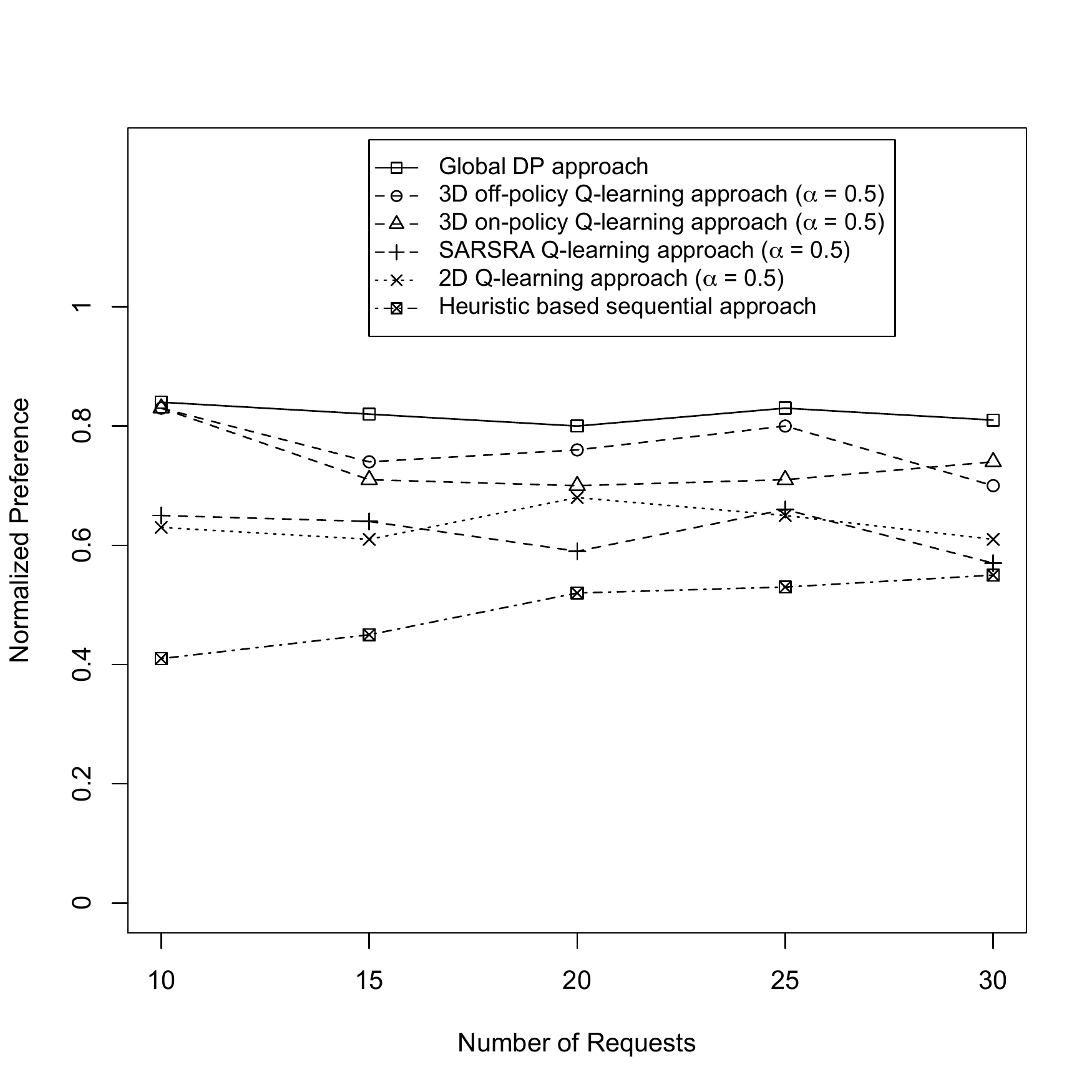}} \hspace*{-.01em}
      \subfloat[]{\includegraphics[width=0.50\textwidth, height=0.5\textwidth]{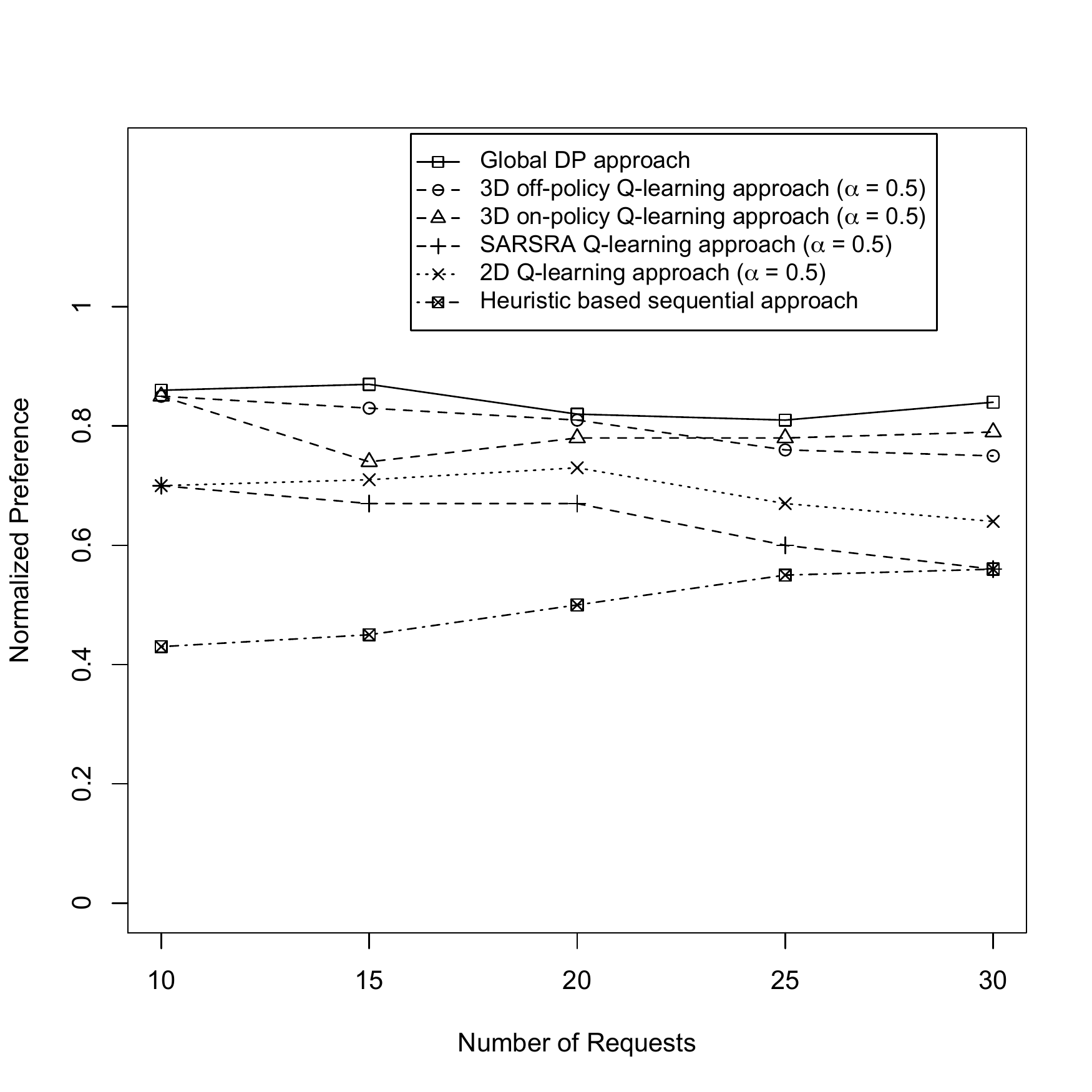}}  \\ [-3ex]
      \subfloat[]{\includegraphics[width=0.50\textwidth, height=0.5\textwidth]{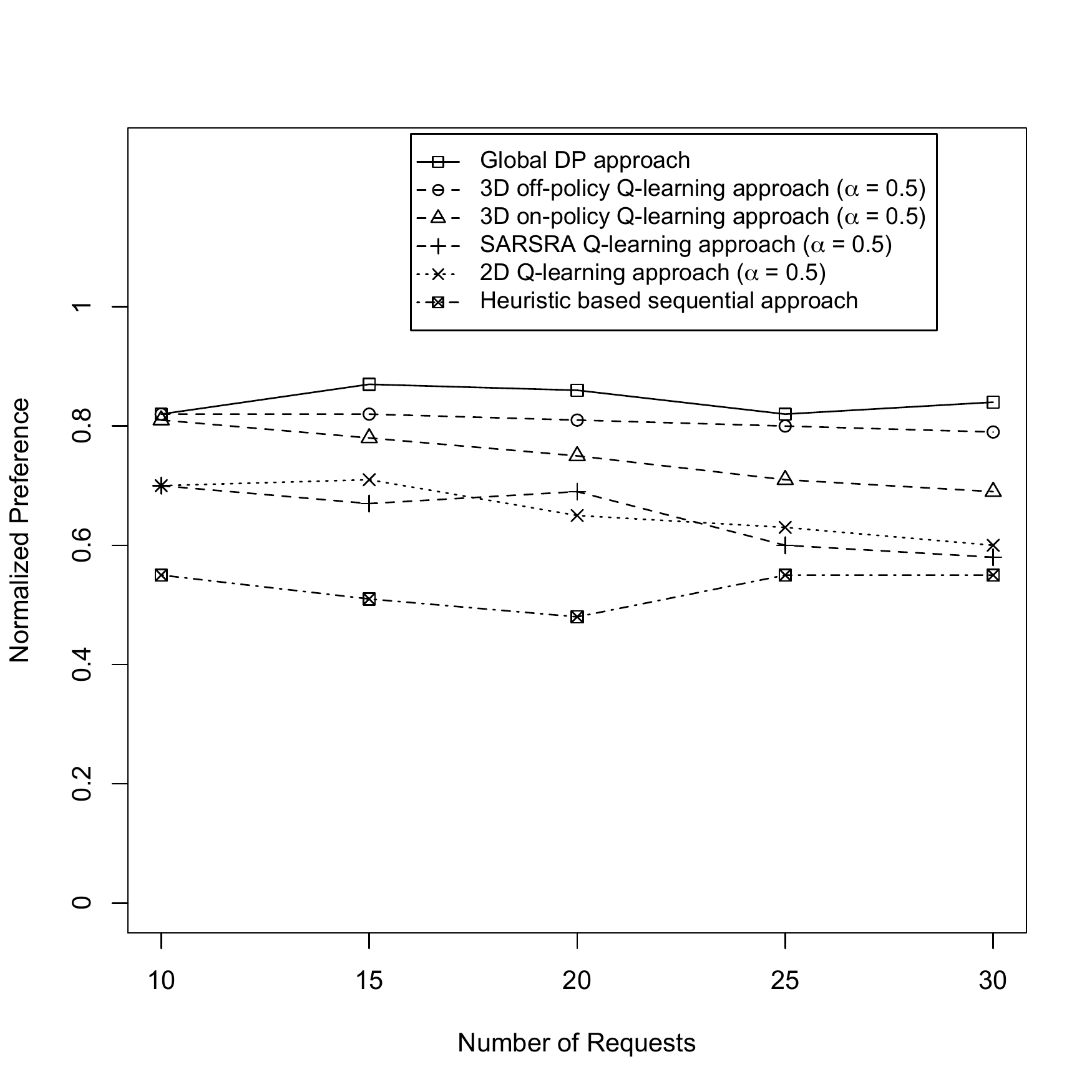}} \hspace*{-.01em}
      \subfloat[]{\includegraphics[width=0.50\textwidth, height=0.5\textwidth]{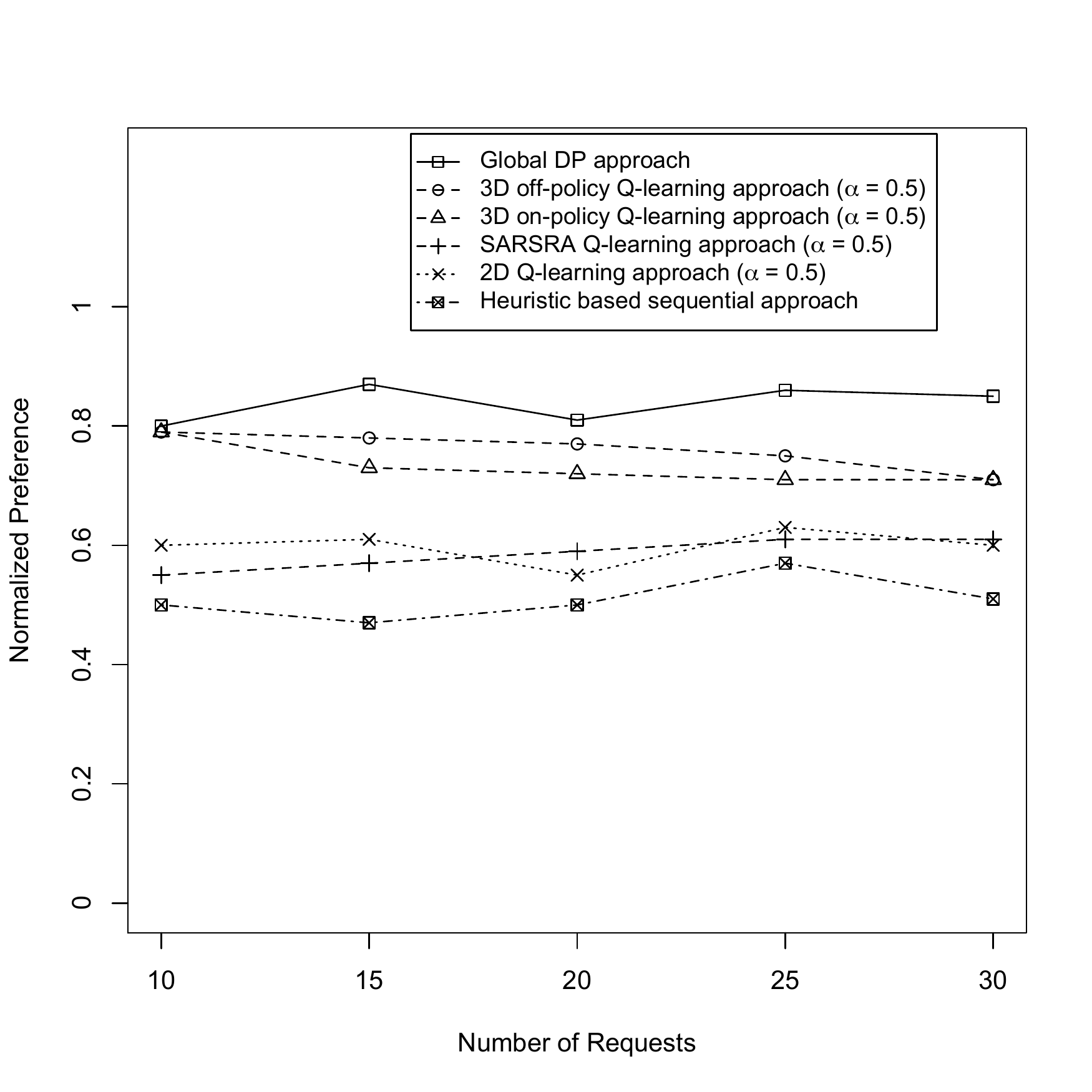}}  \\     
    \caption{\small Accuracy of the DP, on-policy and off-policy Q-learning, SARSA and heuristic approaches in different distributions of requests(a) normal, (b) right-skewed, (c) left-skewed, and  (d) random. \normalsize}
    \label{fig:exp1}
    \vspace{-3mm}
\end{figure}

A higher normalized preference value means higher accuracy and a lower value means lower accuracy. The Figure \ref{fig:exp1} depicts the performances of the proposed approach in different request distributions. In all the distributions (Figure \ref{fig:exp1}), both the off-policy and on-policy 3D Q-learning approach have similar accuracy to the DP based approach. Both approaches have significant higher accuracy than the 2d Q-learning, heuristic-based approaches, and on-policy SARSA approach. The accuracy in the proposed off-policy 3d Q-learning approach is inversely correlated with the number of requests. An increase in the number of requests is followed by a decrease in accuracy for off-policy 3d Q-learning approach irrespective of the type of distribution. It is mainly because an increase in the number of requests results in an increase of transition probabilities in state-action sequences of Q-learning which may cause local optima. There is no significant decrease in the on-policy Q-learning approach for a higher number of request sets except for left-skewed distribution. The possible explanation is that the on-policy Q-learning approach removes redundant state transition probabilities. As there is less number of overlapping requests in left-skewed distribution, the on-policy Q-learning approach behaves similarly to the off-policy 3d Q-learning approach. The accuracy in the heuristic based approach is increased when the number of requests is increased. The accuracy in SARSA is almost similar to the heuristic-based approach. 

The proposed off-policy and on-policy 3D Q-learning approaches demonstrate 35\% higher accuracy than the existing approaches, i.e., SARSA, Q-learning, and Heuristic based approaches in normal, right-skewed, and random distribution. In the case of left-skewed distribution, the accuracy reduces slightly (especially for large number of request sets), but still 15\% higher than the existing approaches. A large portion (around 60\% to 80\%) IaaS requests in the left-skewed distribution is short-term (0-3 months). However, the focus of the paper is long-term compositions. We will focus on achieving the consistent higher accuracy in different distributions in the future work.

\begin{figure}[ht]
\vspace{-8mm}
    \centering
      \includegraphics[width=0.5\textwidth]{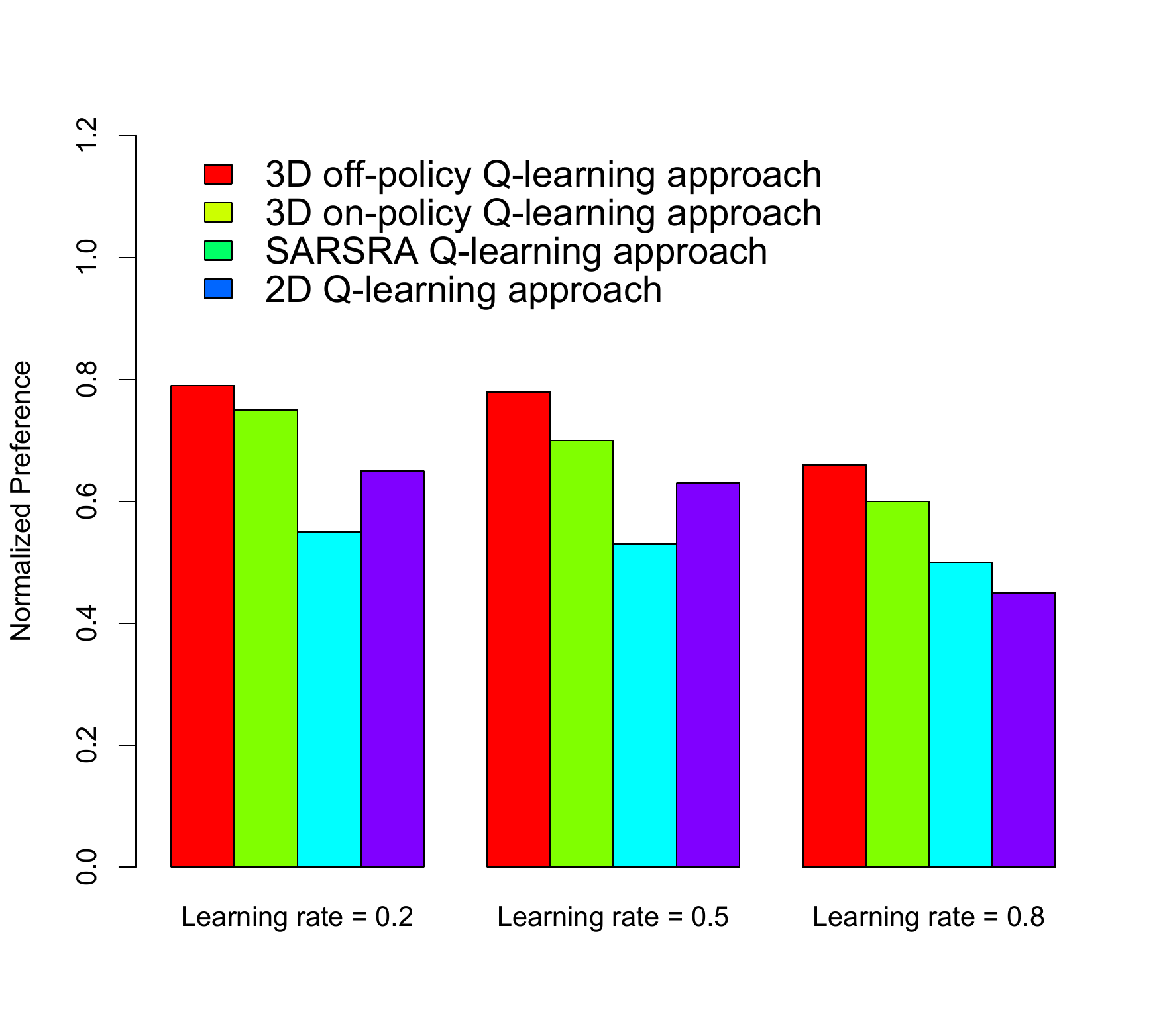} 
      \vspace{-5mm}
    \caption{Preference rankings in different learning rates}
    \vspace{-4mm}
    \label{fig:learning-rate}
\end{figure}

Next, we analyze the effect of learning rate $(\alpha)$ on the accuracy of the proposed framework. We use 3 different learning rates: a) low ($\alpha =0.2$), b) moderate ($\alpha =0.5$), and c) high ($\alpha =0.8$) in the proposed on-policy and off-policy Q-learning approaches in the random distribution set. We randomly choose 10 request sets and the proposed framework is applied to each set separately with the three different learning rates. The averaged accuracy in different learning rate is shown in Fig \ref{fig:learning-rate}. We find that the low and moderate learning rate generates higher and acceptable accuracy for both the on-policy and off-policy 3D Q-learning approaches. However, the higher learning rate reduces the accuracy by around 30\% for the proposed 3d and 2d Q-learning based composition approaches. The key reason is that the higher learning rate overestimates the expected future reward or preference ranking and reduces the state-action exploration process. However, the learning rate has a lower influence on the accuracy of SARSA. As SARSA applies in-policy Q-updates with reduced exploration, different learning rate creates an almost similar number of random explorations which can bring any random update (increase, decrease or remain the same) on the output.

\subsubsection{\textbf{Evaluating Run-time Efficiency in the Composition without History}}

\begin{figure}[ht]
\vspace{-8mm}
    \centering
      \includegraphics[width=0.5\textwidth]{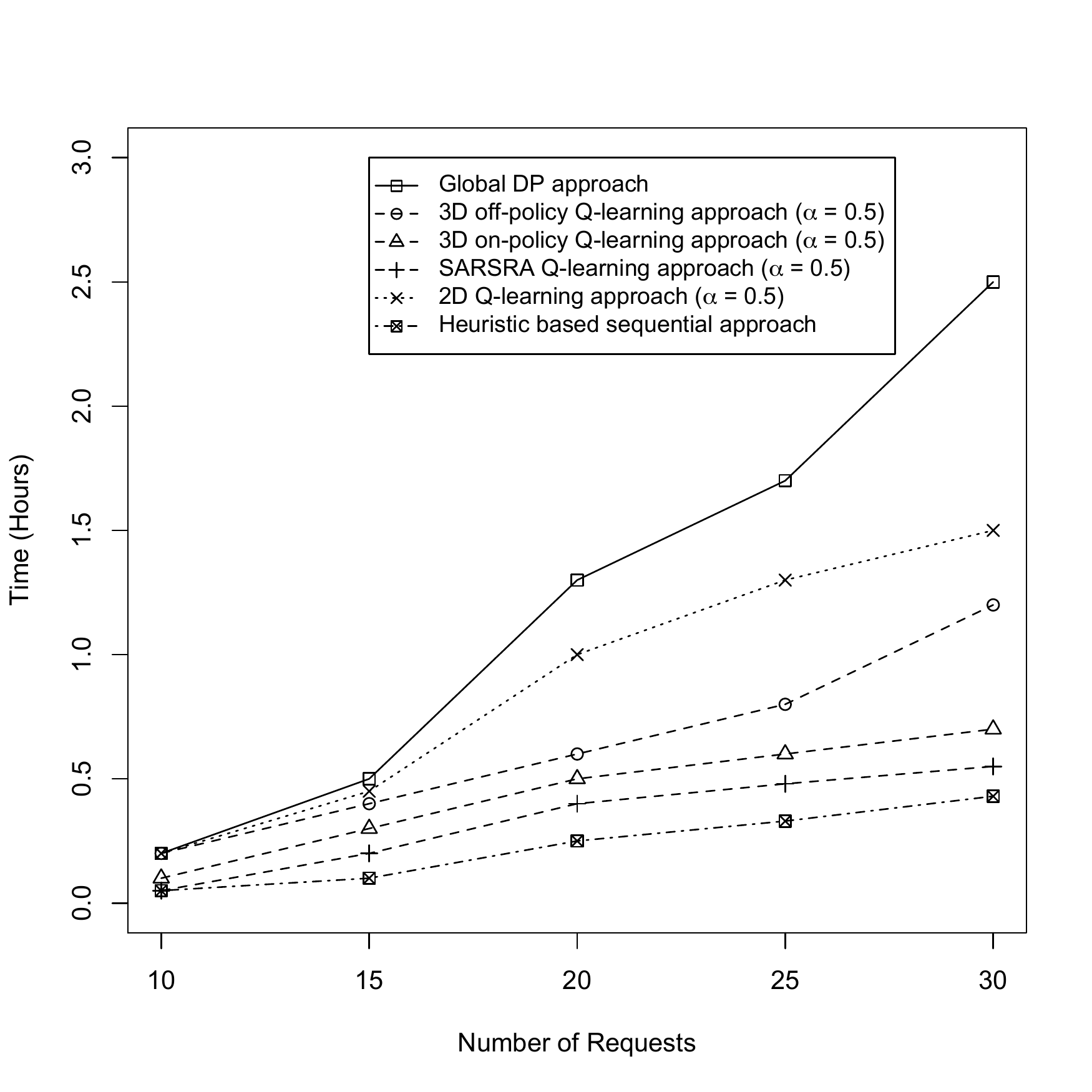}   
    \caption{Runtime efficiency in composition approaches}
    \label{fig:runtimehistoryless}
\end{figure}

We calculate the execution time of each composition for different numbers of incoming requests. We randomly pick 5 requests sets from all the distributions with the same number of request sets. Total 25 compositions are performed on request sets with 10, 15, 20, 25 and 30 incoming requests. The average execution time is described in Figure \ref{fig:runtimehistoryless}. All the composition approaches have similar execution time for a lower number of incoming requests, i.e., sets with 10 and 15 requests. However, for a higher number of requests, the best runtime is generated by the sequential approach and the global DP approach generates the worst case. It defines the applicability of sequential local composition approach over global composition approach for a higher number of requests. We find that the SARSA approach is the quickest among the reinforcement learning approaches due to its on-policy nature. Similarly, the proposed on-policy 3d learning approach runs around 35\% faster than the off-policy 3d q-learning approach and around 65\% faster than the 2d q-learning composition approach. The key reason is the on-policy 3d q-learning approach removes redundant overlapping requests in each episode of its convergence for a composition.

The existing heuristic-based sequential composition approach aggressively reduces the solution space and computation time \cite{sajibicsoc2016}. As a result it is highly effective in real world scenarios where runtime efficiency is more important than accuracy. The proposed approach achieves 35\% higher accuracy using only 5\% increase in computation time than the heuristic-based approach. The impact of this slight increase is negligible in runtime. Moreover, Figure 17 depicts that the proposed approach is scalable in runtime with a higher number of requests set. We conclude that the proposed approach has a higher time efficiency in real world scenarios to achieve the high accuracy.

\subsection{\textbf{Efficiency of the Proposed Policy Reuse Approach with History}}

In the second experiment, we consider that a history of learning experiences are available as a database of learned Q-matrices during a composition. The proposed policy reuse approaches apply the learned policies from a similar set of requests based on the agglomerative clustering approach. All 70 requests are clustered into three sets based on the cophenetic correlation coefficient:

\begin{itemize}
    \item \textit{Set A}: It consists of a highly similar request sets with normalized cophenetic correlation coefficient ranging from (.8 to 1).
    \item \textit{Set B}: It consists of an average similar request sets with normalized cophenetic correlation coefficient ranging from (.6 to .8).
    \item \textit{Set C}: It consists of a lowly similar request sets with normalized cophenetic correlation coefficient ranging from (.3 to .6).
\end{itemize}

\begin{figure}[t!]
 \vspace{-8mm}
      \subfloat[]{\includegraphics[width=0.50\textwidth, height=0.5\textwidth]{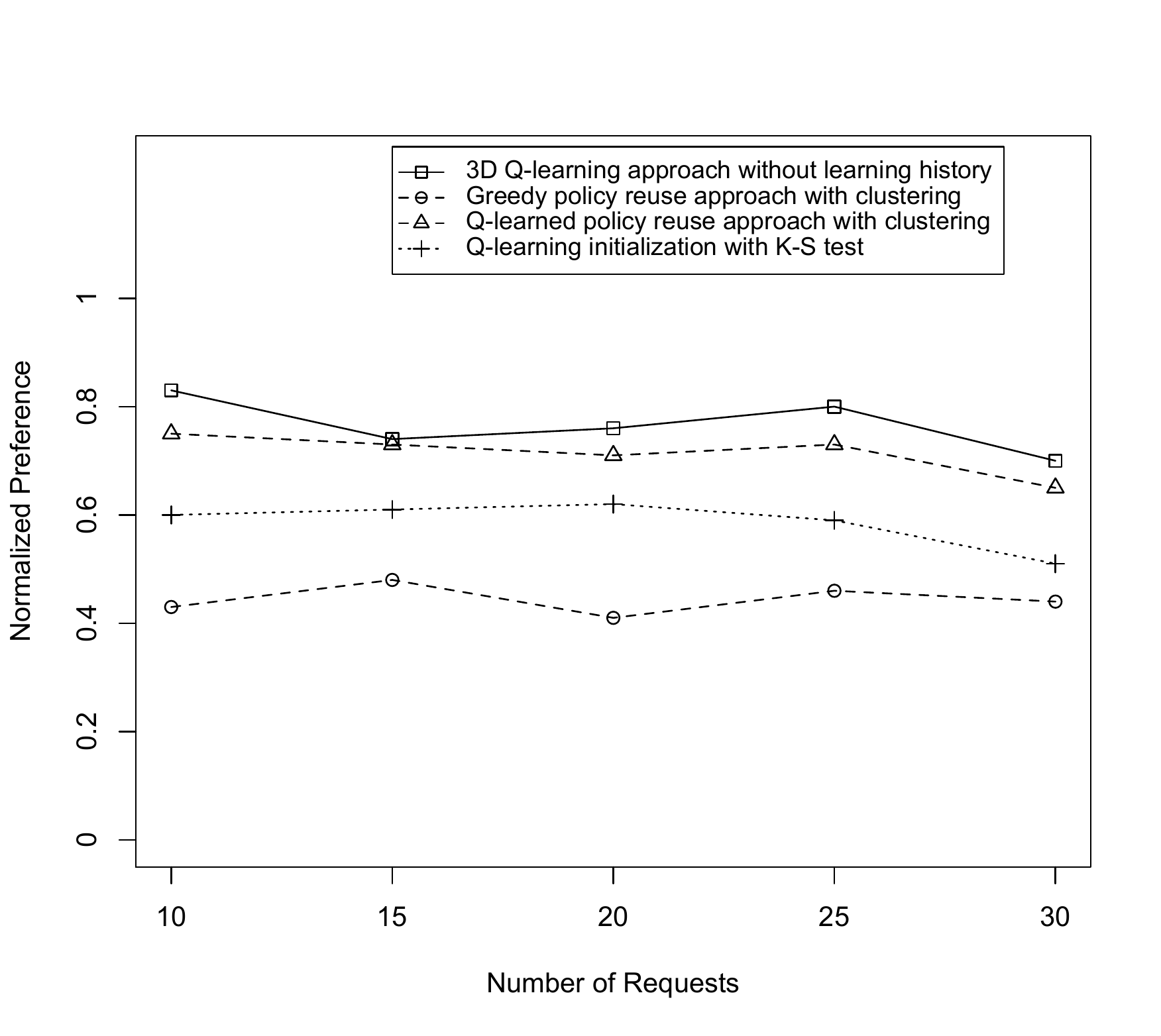}} \hspace*{-.01em}
      \subfloat[]{\includegraphics[width=0.50\textwidth, height=0.5\textwidth]{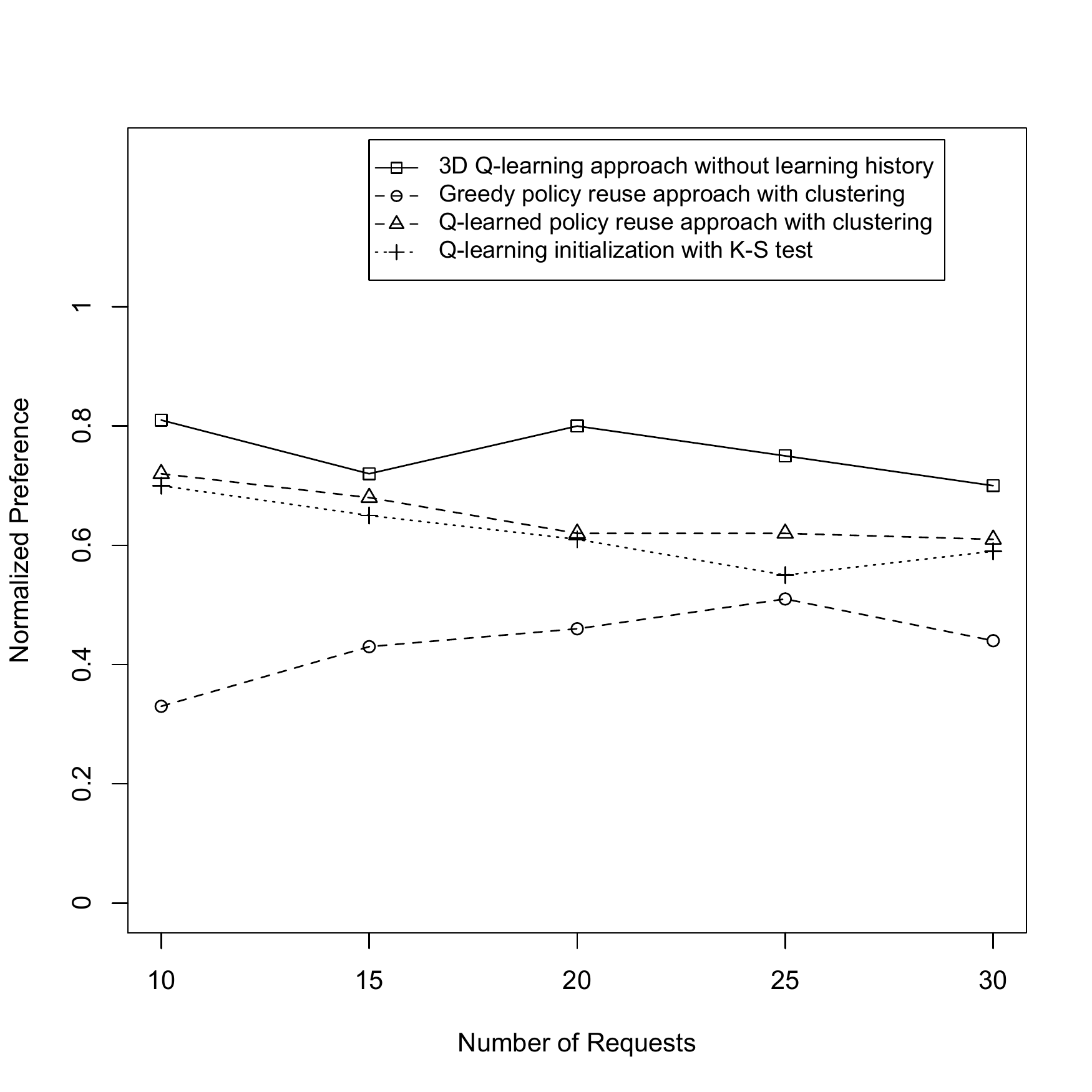}}  \\ [-3ex]
      \subfloat[]{\includegraphics[width=0.50\textwidth, height=0.5\textwidth]{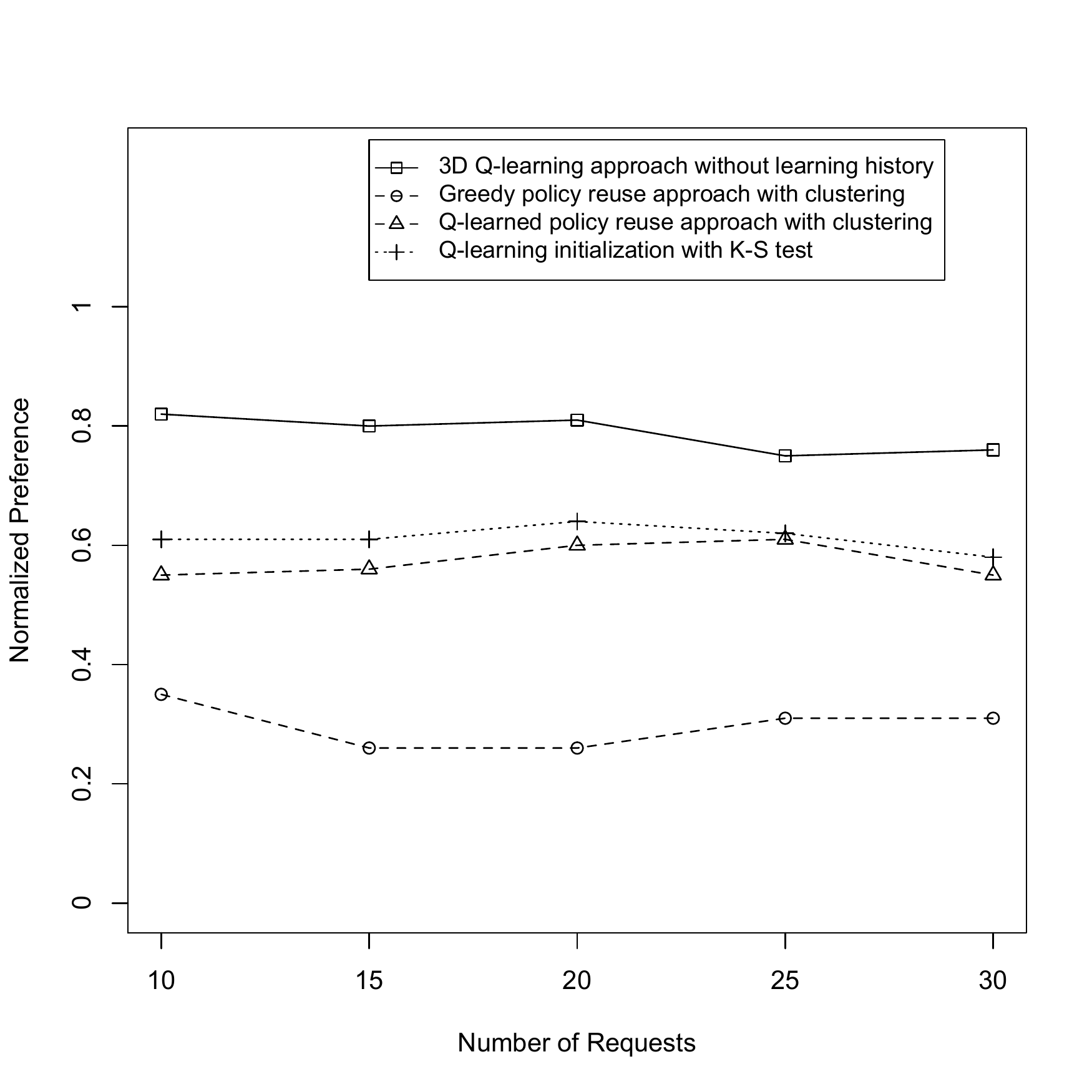}} \hspace*{-.01em}
      \subfloat[]{\includegraphics[width=0.50\textwidth, height=0.5\textwidth]{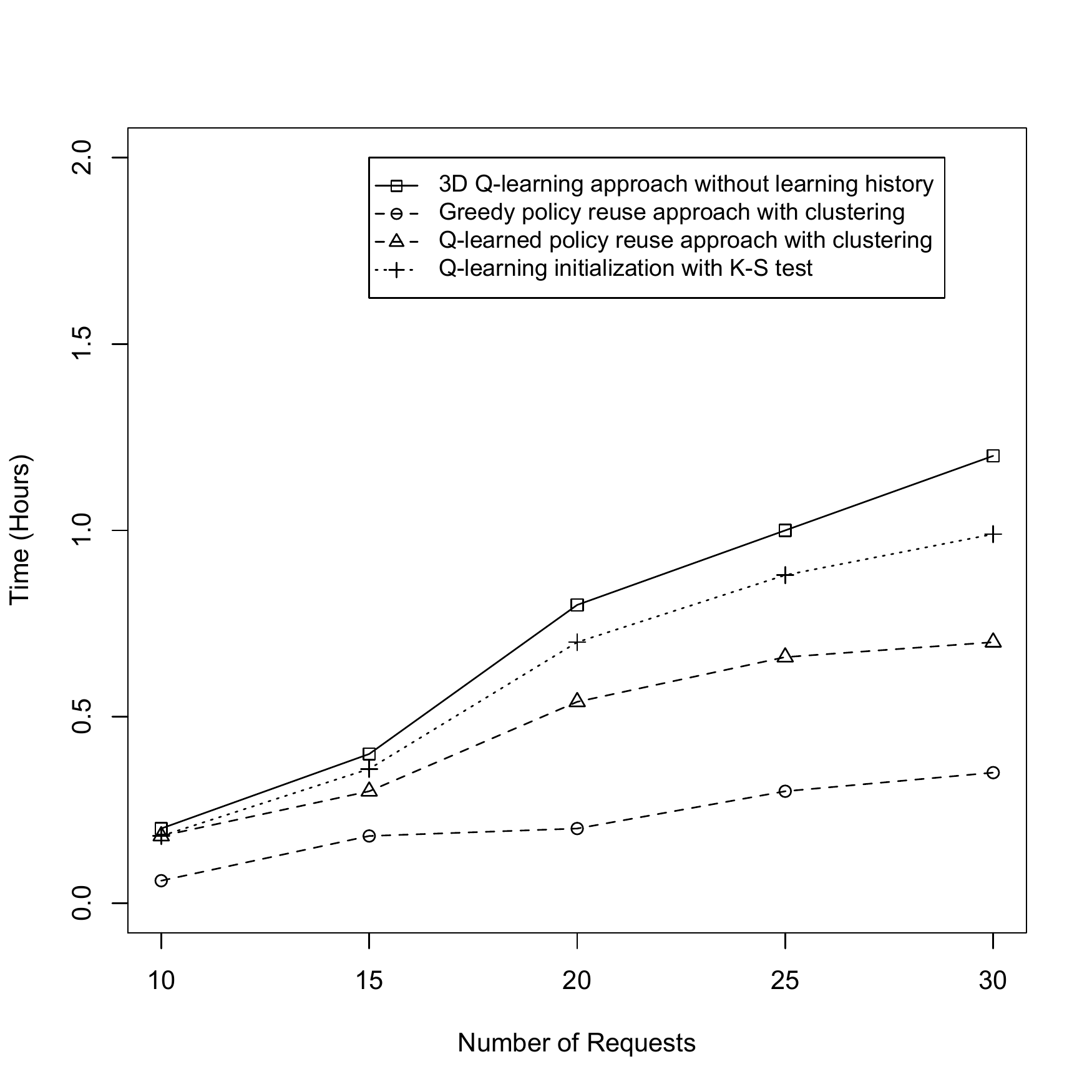}}  \\     
    \caption{\small Accuracy of policy reuse approaches in different distributions, (a) highly similar (Set A), (b) average similar (Set B), (c) lowly similar (Set C), and (d) Average Runtime efficiency in the policy reuse approaches \normalsize}
    \label{fig:exp2}
    \vspace{-3mm}
\end{figure}

Note that, we consider a request set as new if its cophenetic correlation coefficient ranges from 0 to 0.3 when compared with existing request sets. All the requests in set A, B and C are composed individually with the off-policy Q-learning approach and the resulted Q-matrices and policies are stored in a database. Figure \ref{fig:exp2}(a), (b) and (c) depicts the accuracy of the proposed approaches in Set A, B and C. We find that the greedy policy reuse approach performs poorly in every distribution (average accuracy around 35\%). The key reason is that it ignores critical information of request overlapping and the future expected to rank from the current selection. We consider the output from the individual composition without history as a baseline. The efficiency of the policy reuse approach is calculated based on how close output it can produce in comparison to the individual composition without history. According to Figure \ref{fig:exp2}(a), the proposed Q-learned based policy reuse approach performs closely to the output which is generated without history (around 80\% accuracy). However, the accuracy drops to about 60\% in average and lower similar sets (Set B and C) which is not acceptable (Figure \ref{fig:exp2}(b), (c)). The K-S test based approach also has around 60\% accuracy in all the three sets. The K-S test based approach is about 30\% and 5\% less accurate than the proposed Q-learned policy reuse approach in Set A and B respectively (Figure \ref{fig:exp2}(a) and (b)). However, it is 5\% more accurate than the policy based approaches in Set C (Figure \ref{fig:exp2}(c)). The proposed policy reuse approach should only be applied when highly similar request sets are found in the history.

According to existing literature \cite{koenig1993complexity}, the time complexity of 2d Q-learning approach is $O(ep)$, where $e$ is the total number of actions, $e = \sum_{s \epsilon S} a(s)$, and $p$ is the total number of states, $p = |S|$. In our context, the state is dependent on the number of time intervals, i.e., $I$ and the state transitions are laid out on the two dimensional state-space $[I \times I]$. Similarly, the pairwise comparison is performed in each action, i.e., selecting a request set in a composition. Hence, if the number of request set is $N$, the time complexity of 2d Q-learning approach for the IaaS composition is $O(N^{2}I^{2})$. We modified the 2d Q-learning approach for IaaS composition into 3d On-policy learning process in Algorithm 1. As one more dimension is added in the 3d learning, the time complexity of Algorithm 1 is $O(N^{3}I^{3})$. Algorithm 2 applies policy reuse to reduce the state-action transitions by removing the redundant states. As the complexity of the clustering based similarity in Algorithm 2 is $O(N^{2})$, the time complexity of the proposed policy reuse approach is $O(N^{2})$ + $O(N^{2}I^{2})$. Figure \ref{fig:exp2}(d) depicts the average runtime efficiency of the proposed policy reuse approach in the three sets. For a smaller number of sets (10 and 15) the policy reuse approach could not improve the runtime efficiency significantly than the K-S test based approach or composition without history approach. However, the policy reuse approach improves the runtime efficiency significantly (around 40\%) than the Q-learning approach without history for a higher number of requests composition. The runtime of the greedy policy reuse approach is similar to the heuristic based sequential optimization approach.

\begin{figure}[t!]
    \centering
      \includegraphics[width=0.5\textwidth]{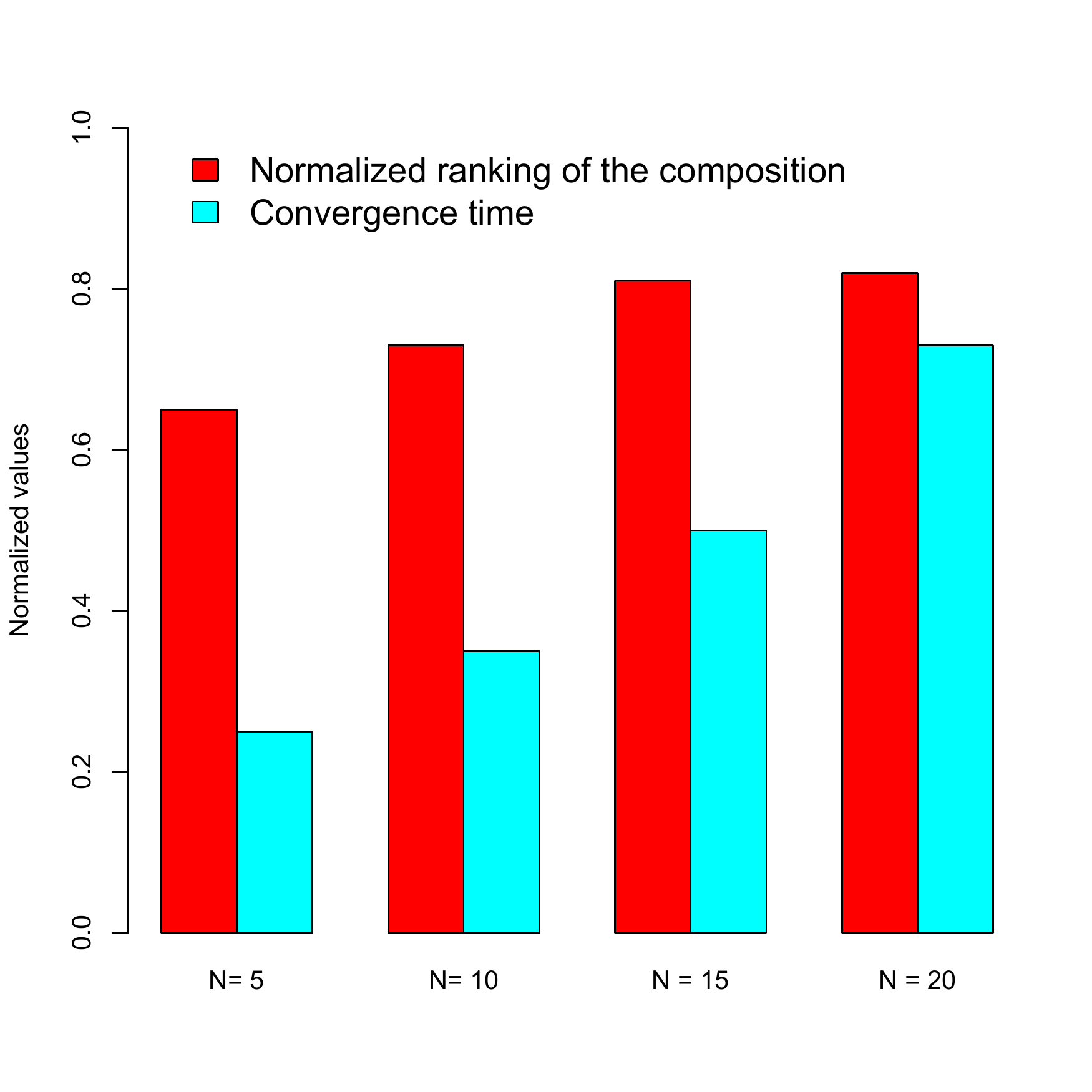}  
      \vspace{-6mm}
    \caption{Effects of different library sizes of Q-matrices in the policy reuse approach}
    \label{fig:librarysize}
\end{figure}

We analyze the effect of the library size, i.e., the number of learned Q-matrices for the proposed policy reuse approach. According to Figure \ref{fig:librarysize}, although a lower library size (i.e., 5) enables a quick convergence for the Q-leaning based policy reuse approach, the resulted accuracy also stays low. The key reason is the exploration method in the learning approach may not receive enough policies from the library of similar policies. Figure \ref{fig:librarysize} depicts that a bigger library size improves the accuracy of the proposed approach. However, the accuracy is not linearly related to the library size. Once, the library size reaches a certain threshold, adding newly learned matrices may not necessarily improve the result. The proposed approach performs a satisfactory level of exploration with the threshold library size and newly added policies may only reiterate over the same explored states. N=15 is the threshold value of library size in Figure \ref{fig:librarysize}. On the other side, a bigger library size also increases the runtime. In Figure \ref{fig:librarysize} the runtime increases at the rate of 20\% if library size is increased by a factor of 5. The threshold library size is dependent on the TempCP-net, the number of intervals and the distribution of the learned Q-matrices. In the future, we will investigate determining the optimal library size of the Q-learned database for the composition framework.

\begin{figure}
    \centering
      \includegraphics[width=0.5\textwidth]{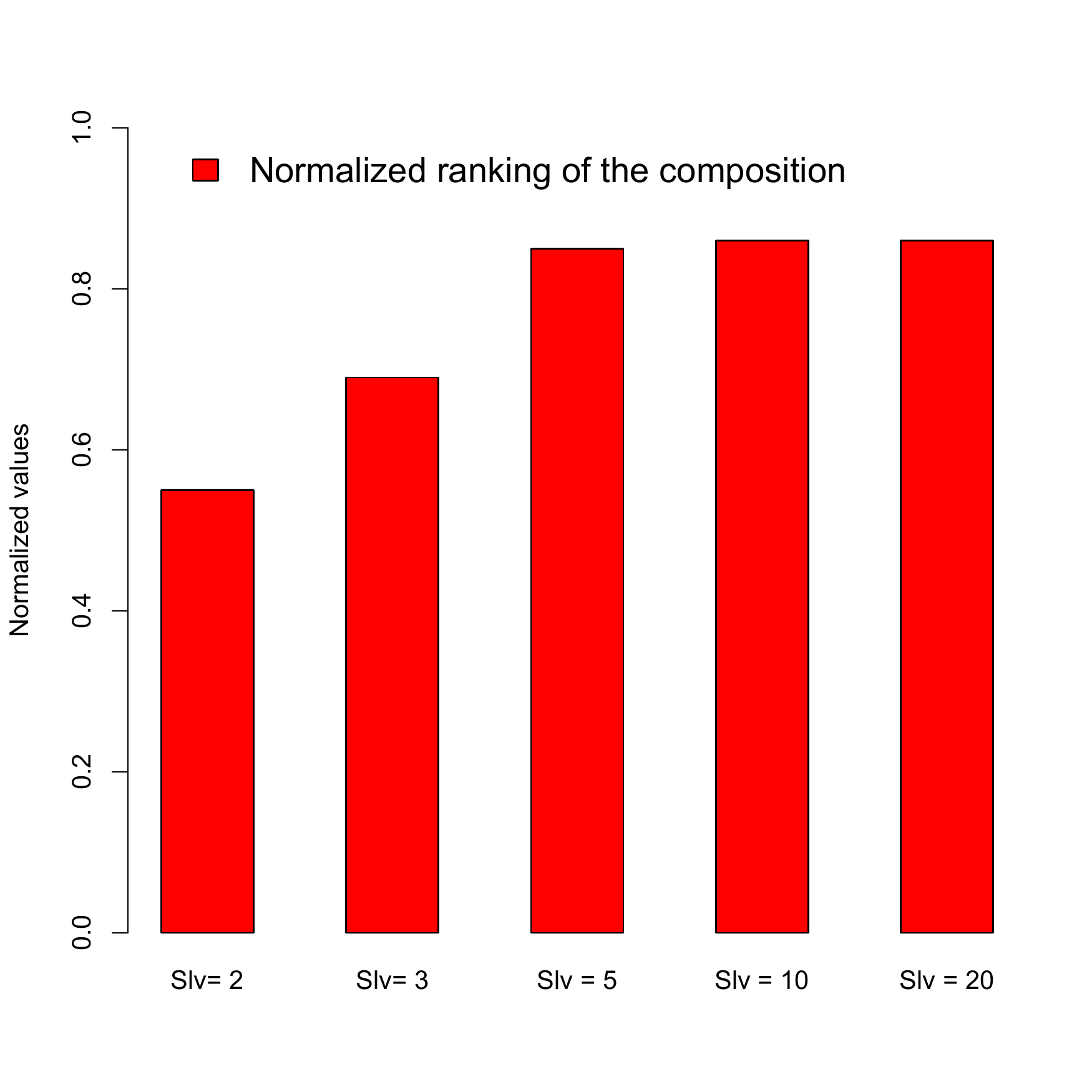}  
    \caption{Effects of different semantic levels in the proposed approach}
    \vspace{-6mm}
    \label{fig:levelsize}
\end{figure}

One of our assumption is that the semantic levels of the provider's qualitative preferences are determined by the providers prior to the composition. The proposed approach determines the optimal qualitative IaaS composition for the provider based on the provided semantic preference table. Note, determining the optimal semantic levels, i.e., value ranges are out of the focus of this paper. We have added a new experiment to depict the effect of different semantic levels $(Slv = [2,3,5,10,20])$ on the accuracy of the proposed approach. The value ranges are uniformly distributed in the experiment. For example, the value range [66,100] is considered ``High'' in 3 semantic levels where the value range [80,100] is considered ``High'' in 5 semantic levels. Figure \ref{fig:levelsize} depicts that a higher semantic levels improves the accuracy of the proposed approach slightly. As the lower semantic levels are coarse-grained, the number of candidate solutions are not effectively filtered in contrast to higher semantic levels. However, after the semantic levels reaches a certain threshold, adding new levels may not necessarily improve the result. In the future, we will investigate determining the optimal semantic levels for the composition framework. 

\section{Related Work}

Cloud service composition has been studied extensively in the existing literature \cite{mistry2018economic}. We identify two types of economic models for cloud service composition - a) quantitative , and b) qualitative composition \cite{mistry2018economic, gavvala2019qos,souri2020hybrid}. We find that most quantitative approaches are mainly considered from a provider's perspective where the objective is to select the optimal set of service requests based on the provider's economic goal such as maximizing profit, revenue, or resource utilization, We also find that qualitative approaches are mainly considered from a consumer's perspective. Composition from a consumer's perspective selects the best cloud service from a set of cloud services based on the consumer's service requirements.


\begin{itemize}
    \item \textit{Quantitative composition from a provider's perspective}: The objective of the quantitative composition is to maximize a provider's economic expectations such as revenue, profit, and reputation by optimizing available resources \cite{cong2020survey}. For example, let us assume an IaaS provider receives requests from two perspective consumers, a bank and a university. The bank requests a VM instance with 8 cores CPU, 1TB memory, 1GB bandwidth, 10ms latency for 1 year and is willing to pay \$500 upfront. The university requests a different VM instance with 16 cores CPU, 1972 CUDA cores GPU, 2TB memory, 500MB bandwidth, 5ms latency for 1 year and is willing to pay \$800 upfront. The provider has four options: a) accept both the requests, b) accept the bank's request, c) accept the university's request, and d) reject both the requests. The economic model of the provider will guide the provider to take the appropriate decision. For example, the economic model of the provider may \textit{suggest} that GPU based services have larger operation costs (due to the higher demand for electricity). As a result, the bank's IaaS request may be \textit{preferred} and \textit{accepted} over the university's IaaS request to \textit{maximize profit}. Note that, if the provider has a \textit{different economic model}, i.e., revenue maximization and reputation building for services with higher throughput, the university's IaaS request may be \textit{preferred} and \textit{accepted} over the bank's IaaS request (the university is paying more upfront that \textit{maximizes revenue}). An economic model is proposed in \cite{goiri2012economic} to maximize resource utilization. The proposed approach assesses the cost of resource utilization from a cloud federation perspective and develops a resource management core for the \textit{short-term} profit maximization. An economic model is proposed to calculate the short-term operation cost of VM provisioning using power ratings of the physical servers in \cite{Thanakornworakij}. The proposed economic model aims at maximizing profit based on right pricing and rightsizing in the Cloud data centre. However, short-term economic benefits may negatively affect the long-term economic gain. A meta-heuristic optimization approach is proposed in \cite{sajibtsc2015} to compose stochastic service requests considering long-term economic gain. However, these proposed quantitative models could not be applied in qualitative composition context as these models do not consider the provider's business strategies or long-term qualitative preferences to accept or reject requests in the composition.
    
    \item \textit{Qualitative composition from a consumer's perspective}: Qualitative economic models are often used to express the preferences of a consumer where there exists incomplete or uncertain information. Qualitative models facilitate the selection and composition of multiple providers when quantitative approaches are not applicable. A qualitative economic model is proposed in \cite{fattah2018cp} to represent a consumer's service preferences. The qualitative model is built using a dependency graph and conditional preference tables. The dependency graph and the conditional preference tables form a Conditional Preference Network (CP-net). A CP-net represents the consumer's preferences qualitatively in a natural and intuitive way. The conditional preference statements for each QoS attribute are expressed using a CPT (Conditional Preference Table). A CP-net based qualitative economic model is presented in \cite{cp1} to select and compose multiple providers based on qualitative preferences of a consumer. The semantic consistency of the CP-net decides the quality of the composition. There are several variations of CP-nets in the existing literature. Consumers are allowed to express the weight or relative importance among attributes using the weighted CP-net (WCP-Net) \cite{wang2012wcp}. The preferences can also be incomplete or inconsistent. A web service selection model using CP-net is proposed in \cite{wang2009web} to select with incomplete or inconsistent user preferences. Probabilities are attached to the CPTs when there exist uncertainties in the qualitative preferences \cite{cornelio2013updates}. A CP-net based user similarity search and conflict removal technique are proposed to remove inconsistency or incompleteness in the user preferences \cite{bohanec2008qualitative}. The qualitative composition approaches from the consumer's perspective could not be applied directly to a provider as multiple long-term requests are composed at the same time according to the qualitative preferences of the provider.   
\end{itemize}

An IaaS provider's preferences are expressed using qualitative economic models. The ordering of the preference outcomes is calculated using an economic variable. Examples of economic variables are cost, reputation, revenue, and so on. The composition depends on the ordering of the preference outcomes. A qualitative economic model usually depends on multiple domain-specific properties \cite{chung2012general}. Several existing studies consider qualitative economic models from the perspective of a provider. A CP-net based composition is proposed to select an optimal set of consumers from a provider's perspective \cite{fattah2018cp}. The proposed approach defines a composability model to determine the inconsistency of the preference composition of the consumers from a provider's perspective. A qualitative economic model for long-term IaaS composition is proposed to capture provider's long-term business strategies using temporal CP-nets \cite{sajibicsoc2016}. A novel sequential optimization method is proposed to select the optimal set of long-term consumer requests. The proposed approach is compared with global dynamic approach and greedy approach which shows that sequential optimization can achieve better time efficiency with acceptable accuracy. However, the proposed method does not include decision variables and business strategies to capture the types of consumers. The decision variables distinguish different types of consumers and enable a natural way to represent qualitative preferences. Decision variables are introduced in \cite{mistry2018long} to capture the provider's qualitative preferences for short-term and long-term requests. The optimal composition is performed using sequential optimization through reinforcement learning following the long-term qualitative economic model.

The sequential local optimization method is a well-known approach in many areas such as operational research, image processing, pattern recognition and so on. Sequential local optimization is usually applicable when the consequences of any given action regarding high-level criteria need to be determined \cite{kimes2004restaurant,chaefer2005modeling}. Many existing approaches use machine learning techniques in a sequential optimization to determine the parameters in state-action sequences \cite{watkins1992q}. As the IaaS composition is dynamic in nature, a model-free learning platform might be more suitable than a model-based learning platform \cite{dorigo2016ant}. Adaptive service composition utilizes reinforcement learning based methods to compose services from a consumer's perspective \cite{wang2020integrating}. Markov Decision Process or MDP is often used to model service composition problem \cite{alizadeh2020reinforcement}. A Q-learning based composition approach is proposed in \cite{lewis2012reinforcement}. The proposed approach models adaptive controllers through a sequential decision process. The optimal policy is constructed using a reward function to approximate the optimal function. A heuristic based optimization approach is proposed without considering historical requests patterns  \cite{sajibicsoc2016}. It creates random sequences of local optimizations. The heuristics treat different types of consumers in a similar manner which may output a non-optimal composition. A reinforcement learning (Q-learning) approach is proposed in \cite{mistry2018long} that learns service provision strategies for different types of consumers without historical information of the past requests. It invalidates the necessity of learning if every new composition needs to be learned from scratch. In this paper, we remove the limitation of the existing approach in \cite{mistry2018long}. We propose a new on-policy 3d Q-learning based IaaS composition framework that has the ability o reuse historical learning experiences or policies using agglomerative clustering.

\section{Conclusion}

We propose a qualitative composition approach for deterministic consumer requests according to the provider's long-term qualitative preferences represented in TempCP-nets. We perform the selection of consumer requests in each temporal segments sequentially. We propose Q-leaning based reinforcement learning approaches to find the best sequence of local optimization or selection of requests. We design an on-policy based three-dimensional Q-learning approach that intuitively removes redundant computations, i.e., state-action transitions using the length of overlapping requests. We propose different policy reuse approaches to perform the composition based on the learned compositions. The similar policy is selected using the cophenetic distance from agglomerative clustering of requests. Experimental results show that the proposed on-policy based 3d Q-learning approach is effective in real world situations. The proposed approach generates more accurate results than other approaches, i.e., 2d Q-learning, SARSA and heuristic based sequential approaches. The accuracy of the proposed on-policy based learning approach closely approximates to the accuracy of the off-policy 3d Q-learning composition approach in significantly reduced runtime. Experimental results also suggest that higher learning rates should not be used in the proposed approach. The proposed on-policy reuse approach produces better accuracy with agglomerative clustering than the greedy and K-S test based reuse approaches in an acceptable runtime. 

One of the key limitations of the proposed work is to determine the optimal library size of Q-matrices which will be addressed in the future work. The proposed framework also does not consider the stochastic arrival of incoming requests and the probabilistic qualitative preferences. We will explore the dynamic qualitative IaaS composition in the future work.

\section{Acknowledgement}

This research was partly made possible by DP160103595 and LE180100158 grants from the Australian Research Council. The statements made herein are solely the responsibility of the authors.

\bibliographystyle{ACM-Reference-Format}
\bibliography{Main.bib}










\end{document}